\documentclass{mn2e}
\usepackage{amsfonts}
\usepackage{amsmath}
\usepackage{graphicx}
\newcommand{\etal}{{et al.~}}
\newcommand{\fFU}{f_{\rm BNC}}

\newcommand{\kmsmpc}{\>{\rm km}\,{\rm s}^{-1}\,{\rm Mpc}^{-1}}
\newcommand{\kms}{\>{\rm km}\,{\rm s}^{-1}}

\newcommand{\Mpc}{\>{\rm Mpc}}
\newcommand{\kpc}{\>{\rm kpc}}
\newcommand{\Msun}{\>{\rm M_{\odot}}}

\newcommand{\beq}{\begin{equation}}
\newcommand{\eeq}{\end{equation}}

\newcommand{\rmd}{{\rm d}}

\newcommand{\calS}{{\cal S}}
\newcommand{\calH}{{\cal H}}
\newcommand{\calC}{{\cal C}}
\newcommand{\calR}{{\cal R}}

\def\gtsima{$\; \buildrel > \over \sim \;$}
\def\ltsima{$\; \buildrel < \over \sim \;$}
\def\prosima{$\; \buildrel \propto \over \sim \;$}
\def\gsim{\lower.7ex\hbox{\gtsima}}
\def\lsim{\lower.7ex\hbox{\ltsima}}
\def\simgt{\lower.7ex\hbox{\gtsima}}
\def\simlt{\lower.7ex\hbox{\ltsima}}
\def\simpr{\lower.7ex\hbox{\prosima}}
\def\la{\lsim}
\def\ga{\gsim}
\def\lta{\la}
\def\gta{\ga}

\newdimen\hssize
\newdimen\hdsize
\begin{document}
%

\title[Brightest Halo Galaxies]
      {Are Brightest Halo Galaxies Central Galaxies?}
\author[Skibba et al.]
       {Ramin A. Skibba$^{1,2}$\thanks{E-mail: rskibba@as.arizona.edu}, 
        Frank C. van den Bosch$^{3}$, 
        Xiaohu Yang$^{4}$, 
        Surhud More$^{1,5}$, \newauthor
        Houjun Mo$^{6}$, 
        Fabio Fontanot$^{1,7}$\\
  $^{1}$Max-Planck-Institute for Astronomy, K\"{o}nigstuhl 17,
	D-69117 Heidelberg, Germany\\
  $^{2}$Steward Observatory, University of Arizona, 933 N. Cherry Avenue, 
        Tucson, AZ 85721, USA\\
  $^{3}$Department of Physics and Astronomy, University of Utah, 
        115 South 1400 East, Salt Lake City, UT 84112-0830, USA\\
  $^{4}$Key Laboratory for Research in Galaxies and Cosmology,
        Shanghai Astronomical Observatory; the Partner Group of MPA;\\ 
        Nandan Road 80, Shanghai 200030, China\\
  $^{5}$Kavli Institute for Cosmological Physics, The University of Chicago, 
        5640 S. Ellis Avenue, Chicago, IL 60637, USA\\
  $^{6}$Department of Astronomy, University of Massachusetts,
        Amherst MA 01003-9305, USA\\
  $^{7}$INAF - Osservatorio Astronomico di Trieste, Trieste, Italy}


\date{}

\pagerange{\pageref{firstpage}--\pageref{lastpage}}
\pubyear{2009}

\maketitle

\label{firstpage}


\begin{abstract}
  It is generally assumed that the central galaxy in a dark matter
  halo, that is, the galaxy with the lowest specific potential energy, is
  also the brightest halo galaxy (BHG), and that it resides at rest at
  the centre of the dark matter potential well. This central galaxy
  paradigm (CGP) is an essential assumption made in various fields of
  astronomical research. In this paper we test the validity of the CGP
  using a large galaxy group catalogue constructed from the Sloan
  Digital Sky Survey. For each group we compute two statistics,
  $\calR$ and $\calS$, which quantify the offsets of the line-of-sight
  velocities and projected positions of brightest group galaxies
  relative to the other group members. By comparing the cumulative
  distributions of $\vert\calR\vert$ and $\vert\calS\vert$ to those
  obtained from detailed mock group catalogues, we rule out the
  null-hypothesis that the CGP is correct. Rather, the data indicate
  that in a non-zero fraction $\fFU(M)$ of all haloes of mass $M$ the
  BHG is not the central galaxy, but instead, a satellite galaxy. 
  In particular, we find that $\fFU$
  increases from $\sim 0.25$ in low mass haloes ($10^{12} h^{-1} \Msun
  \leq M \lta 2 \times 10^{13} h^{-1}\Msun$) to $\sim 0.4$ in massive
  haloes ($M \ga 5 \times 10^{13} h^{-1} \Msun$).  We show that
  these values of $\fFU$ are uncomfortably high compared to
  predictions from halo occupation statistics and from semi-analytical
  models of galaxy formation. We end by discussing various
  implications of a non-zero $\fFU(M)$, with an emphasis on the halo
  masses inferred from satellite kinematics.
\end{abstract}




\section{Introduction}
\label{sec:intro}

According to the current paradigm of galaxy formation, all galaxies
form as a result of gas cooling at the centre of the potential well of
dark matter haloes.  Structures form hierarchically, such that smaller
haloes merge to form larger and more massive haloes.  When a halo and
its `central' galaxy is accreted by a larger halo, it becomes a
subhalo and its galaxy becomes a `satellite' galaxy.  In this 
paradigm, it is assumed that ram-pressure and tidal forces strip
satellite galaxies of their gas reservoir, causing their star
formation to be quenched shortly after having been accreted.  The
central galaxy (i.e., the galaxy with the lowest specific potential
energy), however, continues to accrete new gas, and is also expected to
cannibalize some of its satellites. Consequently, it is generally
assumed that the central galaxy is the most luminous, most massive
galaxy in a dark matter host halo, and that it resides at rest at the
centre of the halo's potential well.  Following van den Bosch et
al. (2005; hereafter vdB05), we will refer to this as the `Central
Galaxy Paradigm' (CGP).

There are numerous areas of astronomy in which the validity of the CGP
is an essential assumption, although this is rarely enunciated.
Examples are various techniques to measure halo masses, such as
satellite kinematics (e.g., McKay \etal 2002; van den Bosch et
al. 2004; More \etal 2009), weak lensing (e.g., Mandelbaum et
al. 2006; Johnston \etal 2007; Cacciato \etal 2009; Sheldon \etal
2009b), and strong lensing (e.g., Kochanek 1995; Cohn \etal 2001;
Koopmans \& Treu 2003; Rusin \etal 2003).  In addition, the CGP also
features in halo occupation modelling, where assumptions have to be
made regarding the distribution of galaxies within dark matter haloes
in order to compute the galaxy-galaxy correlation function on small
scales (e.g., Scoccimarro \etal 2001; Sheth \etal 2001; Yang et
al. 2003; Zehavi et al. 2005; Zheng \etal 2005; Cooray 2005; van den
Bosch \etal 2007; Tinker \etal 2008), and in algorithms developed to
identify galaxy groups and clusters in photometric or spectroscopic
redshift surveys (e.g.,  Yang \etal 2005a, 2007; Berlind \etal 2006;
Koester \etal 2007).

Whether central galaxies actually comprise a special population has
been debated for many years.  However, recent analyses of galaxies in
groups and clusters (e.g., Weinmann \etal 2006; Skibba \etal 2007; von
der Linden \etal 2007; van den Bosch \etal 2008; Pasquali et
al. 2009, 2010; Skibba 2009; Hansen \etal 2009) and halo model analyses
of galaxy clustering (e.g., Skibba \etal 2006; Cooray 2006; van den
Bosch \etal 2007; Skibba \& Sheth 2009) have explicitly shown that
central galaxies are indeed a distinct population and exhibit
different properties (e.g., colour, star formation activity, AGN
activity, morphology, stellar population properties) than satellite
galaxies of the same stellar mass, and with different dependencies on
the mass of their host halo. In addition, these studies have shown
that central galaxy properties are strongly correlated with halo mass,
while those of satellite galaxies only reveal a very weak dependence
on the mass of the halo in which they orbit.  However, it is important
to realize that in virtually all of these studies, central galaxies are
{\it assumed} to the the brightest (or most massive) halo galaxies. If
this aspect of the CGP is not correct for a non-negligible fraction of
all haloes, the differences between centrals and satellites found by
these studies will have to be considered lower limits.

The validity of the CGP has been investigated by a number of
authors. In particular, several recent, observational studies have
shown that although most brightest halo galaxies (hereafter BHGs) are
nearly at rest near the centroid of their group or cluster, or at the
peak of the cluster X-ray emission, some of them are not (e.g., Beers
\& Geller 1983; Malumuth \etal 1992; Bird 1994; Postman \& Lauer 1995;
Zabludoff \& Mulchaey 1998; Oegerle \& Hill 2001; Yoshikawa \etal
2003; Lin \& Mohr 2004; von der Linden \etal 2007; Bildfell \etal 2008; 
Hwang \& Lee 2008; Sanderson et al. 2009; Coziol \etal 2009).  
These studies have focused on either cD galaxies in clusters, or on 
brightest cluster galaxies (BCGs) in general.  It has been argued 
that most cD galaxies form a subpopulation of BCGs (e.g., Bernstein \& 
Bhavsar 2001; Coziol \etal 2009) and may have grown by ``cannibalizing'' 
smaller neighboring galaxies.

In an early study, Beers \& Geller (1983) analyzed the spatial
distribution of bright galaxies in 56 rich clusters and argue that cD
galaxies tend to lie at local density peaks but not necessarily at the
bottom of the potential well of the whole cluster.  Oegerle \& Hill
(2001) found that, out of their sample of 25 Abell clusters, the cD
galaxies of four of them have significant peculiar velocities relative
to the cluster velocity.  More recent studies have analyzed the
positions and velocities of BCGs.  For example, in a study of 833 SDSS
clusters, von der Linden \etal (2007) found that 21 BCGs in their
sample lie further than 1 Mpc away from the mean galaxy in the
cluster.  Hwang \& Lee (2008) found that the BCGs of two clusters out
of a sample of 24 have significantly offset velocities and positions;
these two clusters also appear to be in dynamical equilibrium.
Recently, Coziol \etal (2009), using a sample of 452 Abell clusters
selected for the likely presence of a dominant galaxy, estimated that
the BCGs have a median peculiar velocity of 32\% of their host
clusters' radial velocity dispersion.

We emphasize that these results based on large clusters do not
necessarily hold for less massive haloes. After all, in galaxy
clusters, the ratio $L_2/L_1$ of the luminosities of the two brightest
galaxies tends to be much smaller in a Milky Way (MW)-sized halo, on
average (e.g., van den Bosch et al. 2007). For example, for the halo
hosting the MW it is the ratio of the luminosities of the Large
Magellanic Cloud and the MW, which is $\sim 0.1$ (van den Bergh 1999),
while this ratio is much closer to unity in Virgo, Coma and other
nearby clusters (Postman \& Lauer 1995). Consequently, it is only
natural to expect that the CGP is more likely to be valid for low mass
haloes than for massive cluster-sized haloes. Nevertheless, vdB05 used
a large sample of 3473 galaxy groups from the group catalogue of Yang
et al. (2005a), and conclusively falsified the assumption that the
brightest group galaxy is always at rest at the centre of the
potential well. In this paper we extend the analysis of vdB05 using a
larger, more accurate group catalogue, and focusing on different
aspects of the CGP as a function of halo mass.  In particular, we
separately test two aspects of the CGP, namely, (i) central galaxies
reside at rest at the centre of their host halo's potential well, and
(ii) central galaxies are the brightest, most massive galaxies in
their host haloes. We do so by comparing three hypotheses:
\begin{itemize}
\item $\calH_0$: Our null hypothesis is that the CGP is correct:
  central galaxies are always the brightest objects in their haloes and
  are at rest at the centre of the potential well.
\item $\calH_1$: Central galaxies are the brightest objects in their
  haloes, but they have a velocity and spatial offset with respect to
  the centre of the potential well, such that the systems still obey
  the Jeans equations (i.e., they are still in dynamical equilibrium).
  We will specify the amount of offset via a velocity bias parameter,
  $b_{\rm vel}$, to be defined in Section~\ref{model}.
\item $\calH_2$: Central galaxies reside at rest at the centre of the
  potential well, but they are not the brightest objects in a fraction
  $\fFU$ (for `Brightest-Not-Central') of all dark matter haloes.
\end{itemize}
In order to avoid confusion, throughout this paper we use the term
`central galaxy' to refer to the galaxy with the lowest specific
potential energy.  The central galaxy is the BHG in $\calH_0$ and
$\calH_1$, but not in $\calH_2$, and its location coincides with the
centre of the halo's potential well in $\calH_0$ and $\calH_2$, but
not in $\calH_1$.

We base our study on the galaxy group catalogue of Yang \etal (2007),
extracted from the Sloan Digital Sky Survey (SDSS; York \etal 2000)
Data Release 4 (DR4; Adelman-McCarthy \etal 2006).  We analyze both
the positions and velocities of BHGs relative to those of the other
member galaxies, and compare the results to mock group catalogues that
correspond to one of our three hypotheses.  The large SDSS group
catalogue, with accurate halo mass estimates, allows us to falsify
$\calH_0$ and to quantify the degree to which $\calH_1$ and $\calH_2$
are valid (i.e., to constrain the values of $b_{\rm vel}$ and $\fFU$).

This paper is organized as follows.  In Section~\ref{RparamSect}, we
present two statistics that quantify the offsets of BHGs, and which
can be used to assess deviations from the central galaxy paradigm.  We
describe the SDSS group catalogue in Section~\ref{grpcat} and the
construction of mock group catalogues in Section~\ref{sec:mocks}.  In
Section~\ref{sec:res}, we compare these mock catalogues to the data,
in order to test our three hypotheses.  We find that both $\calH_0$
and $\calH_1$ can be ruled out, but that $\calH_2$ yields results in
agreement with the data as long as $0.25 \lta \fFU \lta 0.4$, with a
weak dependence on halo mass.  In Section~\ref{sec:disc}, we discuss
our results and compare them to predictions from halo occupation
statistics and from two semi-analytic models of galaxy formation.  We
also discuss the implications for studies of satellite kinematics.
Finally, we end the paper with our conclusions and a discussion of
additional implications of our results.

Throughout this paper, we adopt a flat $\Lambda$CDM cosmology with
$\Omega_m=0.238$, $\Omega_\Lambda=1-\Omega_m$, $n=0.951$,
$\sigma_8=0.744$, and we express units that depend on the Hubble
constant in terms of $h \equiv H_0/100\kmsmpc$.  In addition, we use
`log' as shorthand for the 10-based logarithm.

\section{Phase-Space Statistics of Central Galaxies}
\label{RparamSect}

In order to test the CGP (i.e., hypothesis $\calH_0$), we use the
line-of-sight velocities of galaxies obtained from their redshifts.
In what follows, $v_{\rm BHG}$ refers to the line-of-sight velocity of
the BHG, and $v_{{\rm sat},i}$ refers to the line-of-sight velocity of
the $i^{\rm th}$ satellite galaxy.  We define the difference $\Delta
V={\bar v}_{\rm sat}-v_{\rm BHG}$ between the {\it mean} velocity of
the satellite galaxies and that of the BHG.  If the CGP is correct and
$v_{{\rm sat},i}$ follows a Gaussian distribution with velocity
dispersion $\sigma_{\rm sat}$, then the probability that a halo with
$N_{\rm sat}$ satellite galaxies has a value of $\Delta V$ is given by
\begin{equation}
 P(\Delta V){\rm d}\Delta V \,=\, \frac{1}{\sqrt{2\pi}\sigma} \,
    \exp\left[-\frac{(\Delta V)^2}{2\sigma^2}\right] \,\rmd\Delta V
\end{equation}
where $\sigma=\sigma_{\rm sat}/\sqrt{N_{\rm sat}}$.  Therefore,
in principle, one could define the parameter
\begin{equation}
R = \frac{\sqrt{N_{\rm sat}}({\bar v}_{\rm sat}-v_{\rm BHG})}{\sigma_{\rm sat}} ,
\end{equation}
and test the CGP by checking whether $R$ follows a normal distribution
with zero mean and unit variance.  However, the velocity dispersion
$\sigma_{\rm sat}$ is generally unknown, and instead we must use
its unbiased estimator
\begin{equation}\label{sigmavsat}
 \hat{\sigma}_{\rm sat} = \sqrt{{1\over N_{\rm sat}-1} 
\sum_{i=1}^{N_{\rm sat}} (v_{{\rm sat},i}-{\bar v}_{\rm sat})^2 }\,.
\end{equation}
Following vdB05, we use the following modified parameter as an
indicator of the offset between BHGs and satellite galaxies:
\begin{equation}\label{Rparam}
\calR = \frac{\sqrt{N_{\rm sat}}({\bar v}_{\rm sat}-v_{\rm BHG})}
{\hat{\sigma}_{\rm sat}}\,.
\end{equation}
This parameter is similar to the relative peculiar velocities of
brightest cluster galaxies, $|v_{\rm pec}| / \sigma_{\rm cluster}$,
used in related studies (e.g., Malumuth \etal 1992; Coziol
\etal 2009).  If the null-hypothesis of the CGP is correct, $\calR$
should follow a Student $t$-distribution with $\nu=N_{\rm sat}-1$
degrees of freedom.  Note that $P_{\nu}(\calR)$ approaches a normal
distribution with zero mean and unit variance in the limit $N_{\rm
  sat} \rightarrow \infty$.

In practice, although the galaxy group finder (Yang \etal 2005a, 2007)
has been thoroughly tested with mock SDSS catalogues to reliably
identify galaxies residing in the same dark matter halo, it is not
perfect.  In particular, because of redshift errors and redshift-space
distortions, the group finder inevitably selects some interlopers
(galaxies that are not associated with the same halo).  In addition,
the SDSS suffers from various incompleteness effects.  If the actual
BHG is missed or misidentified, $\calR$ will be measured with respect
to a satellite galaxy, and $|\calR|$ will tend to be overestimated.
The presence of interlopers and incompleteness effects tend to create
excessive wings in the $\calR$ distribution, and therefore a direct
comparison with the Student $t$-distribution cannot be made.  To
circumvent these problems, we compare the $\calR$-distributions
obtained from galaxy groups identified in the SDSS to those obtained
from groups identified in mock galaxy redshift surveys
(Section~\ref{sec:mocks}), which suffer from interlopers and
incompleteness to the same extent as the real data.

We also investigate the spatial offsets of BHGs in this paper, and to
do so we introduce the following parameter, analogous to the parameter
$\calR$, that quantifies the spatial separation between BHGs and
satellite galaxies, using their projected angular separations
perpendicular to the line-of-sight:
\begin{equation}
\calS = \frac{ \sqrt{N_{\rm sat}} ({\bar r}_{p,{\rm sat}} - r_{p,{\rm BHG}}) }
  {\hat{\sigma}_{r_{p,{\rm sat}}} }
 \label{Sparam}
\end{equation}
where ${\bar r}_{p,{\rm sat}}$ is the mean projected position of
satellite galaxies in a group, in terms of the galaxies' mean right
ascensions and declinations, and
\begin{equation}
\hat{\sigma}_{r_{p,{\rm sat}}} = \sqrt{{1 \over N_{\rm sat}-1} 
\sum_{i=1}^{N_{\rm sat}} (r_{p,{\rm sat},i}-{\bar r}_{p,{\rm sat}})^2 } .
\end{equation}
Both ${\bar r}_{p,{\rm sat}} - r_{p,{\rm BHG}}$ and
$\hat{\sigma}_{r_{p,{\rm sat}}}$ are expressed in $h^{-1} \kpc$ in the
computation of $\calS$.

For the mean and standard deviation ${\bar v}_{\rm sat}$ and
$\hat{\sigma}_{\rm sat}$ in the $\calR$ parameter (Eqn.~\ref{Rparam}),
and ${\bar r}_{p,{\rm sat}}$ and $\sigma_{r_{p,{\rm sat}}}$ in the
$\calS$ parameter (Eqn.~\ref{Sparam}), we have tested that other
estimators for these quantities, such as the biweight estimator and
the gapper (Beers \etal 1990), have an insignificant effect on our
results.

Note that, with these definitions, the velocity offset $\calR$ and 
spatial offset $\calS$ are only valid for groups with three or more
members, and are not defined for galaxy pairs or isolated galaxies.

\section{Application to the SDSS}
\label{grpcat}

We first describe the SDSS galaxy group catalogue in
Section~\ref{grpcatintro} and then the subset of galaxy groups used in
our analysis in Section~\ref{grpsubset}.

\subsection{Galaxy Group Catalogue}
\label{grpcatintro}

The analysis presented in this paper is based on the SDSS galaxy group
catalogue of Yang \etal (2007; hereafter Y07), which is constructed by
applying the halo-based group finder of Yang \etal (2005a) to the New
York University Value-Added Galaxy Catalog (NYU-VAGC; see Blanton
\etal 2005), which is based on the SDSS DR4 (Adelman-McCarthy \etal
2006).  From this catalogue Y07 selected all galaxies in the Main Galaxy
Sample (Strauss \etal 2002) with an extinction corrected apparent
magnitude brighter than $m_r = 18$, with redshifts in the range $0.01
\leq z \leq 0.20$ and with a redshift completeness $\calC_z > 0.7$.

This sample of galaxies is used to construct three group samples:
sample I, which only uses the 362356 galaxies with measured redshifts
from the SDSS, sample II which also includes 7091 galaxies with SDSS
photometry but with redshifts taken from alternative
surveys\footnote{These redshifts are taken from the 2dFGRS
(Colless et al. 2001), \textit{IRAS} PSC$z$ (Saunders et al. 2000), or
RC3 (de Vaucouleurs et al. 1991). See Blanton et al. (2005) for
details.}, and sample III which includes an additional 38672 galaxies
that lack a redshift due to fiber collisions, but which we assign the
redshift of its nearest neighbor.  The present analysis is based on
the galaxies in sample II with $m_r < 17.77$, which consists of 344010
galaxies.

All the magnitudes and colours of the galaxies are Petrosian, and they
have been corrected for Galactic extinction (Schlegel, Finkbeiner \&
Davis 1998), and have been $k$-corrected and evolution-corrected to
$z=0.1$ (Blanton \& Roweis 2007).  Stellar masses for all galaxies are
computed using the relations between stellar mass-to-light ratio and
colour of Bell \etal (2003).

The geometry of the SDSS used for the group catalogue is defined as the
region on the sky that satisfies the redshift completeness criterion.
To account for the effects of the survey edges, Y07 used the SDSS DR4
survey mask with mock galaxy redshift surveys to estimate the fraction
of ``missing'' members within the halo radius for each group.  The
group luminosities and masses are corrected for this fraction, and
groups missing $40\%$ or more of their members were excluded, which
removes only $1.6\%$ of all groups.

As described in Y07, the majority of the groups in our catalogue have
two estimates of their dark matter halo mass: one based on the ranking
of its total characteristic luminosity, and the other based on the
ranking of its total characteristic stellar mass, both determined from
group galaxies more luminous than $M_r - 5 \log h=-19.5$.  As shown in
Y07, both halo masses agree very well with each other, with an average
scatter that decreases from $\sim0.1$ dex at the low mass end to
$\sim0.05$ at the massive end.  In this paper we adopt the group
masses based on the stellar mass ranking, but we have checked that the
luminosity ranking gives results that are almost indistinguishable.
The stellar mass based group masses are available for a total of
215493 groups in our sample, which host a total of 277838 galaxies.
This implies that a total of 66172 galaxies have been assigned to a
group for which no reliable mass estimate is available (but see Yang,
Mo \& van den Bosch 2009a).

\subsection{Galaxy Groups used in this Paper}
\label{grpsubset}

In what follows we restrict our analyses to the 7234 galaxy groups in
the sample II catalogue with three or more members, with $50\kms \leq
\hat{\sigma}_{\rm sat} \leq 1000\kms$, and with reliable group masses
greater than $10^{12}\,h^{-1}\,M_\odot$.

Because of the finite thickness of the spectroscopic fibers used, the
SDSS suffers from incompleteness due to fiber collisions.  No two
fibers on the same SDSS plate can be closer than 55 arcsec.  Although
this fiber collision constraint is partially alleviated by the fact
that neighboring plates have overlap regions, $\sim 7$ percent of all
galaxies eligible for spectroscopy do not have a measured redshift.
Since fiber collisions are more frequent in regions of high
(projected) density, they are more likely to occur in richer groups,
thus causing a systematic bias that may need to be accounted for.
Although Sample III tries to correct for this incompleteness by
assigning galaxies that lack a redshift due to fiber collisions the
redshift of its nearest neighbor, Zehavi \etal (2002) have shown that
in roughly 40 percent of cases, the redshift thus assigned carries a
large error.  Hence, although Sample II is more incomplete than Sample
III, there is less of a risk of interlopers, and the redshifts are
accurate, which is necessary for the analysis with galaxy
line-of-sight velocities.  However, if the true brightest galaxy in a
group is missed due to a fiber collision, the velocity and spatial
offsets $\calR$ and $\calS$ will be estimated relative to a satellite
galaxy, and will tend to be overestimated, resulting in a stronger
signal.  To avoid this problem, we exclude groups in Sample II in
which the brightest galaxy is not also a brightest group galaxy in
Sample III.

This results in a sample of 6760 groups with $N_{\rm gal}\geq3$,
excluding approximately 7\% of the galaxy groups. This constitutes our
fiducial galaxy group sample.\footnote{Using the additional criterion
that the group masses in Samples II and III are within 0.3 dex of each
other excluded another 7\%, but yielded results that were
indistinguishable from those based on our fiducial sample.}
To be clear, our fiducial sample simply consists of a subset of Sample
II groups.  We emphasize that Sample III groups are not used in our
analysis; they are only used to remove those groups from Sample II
that may have been affected by fiber collisions.

\begin{figure}
\includegraphics[width=\hsize]{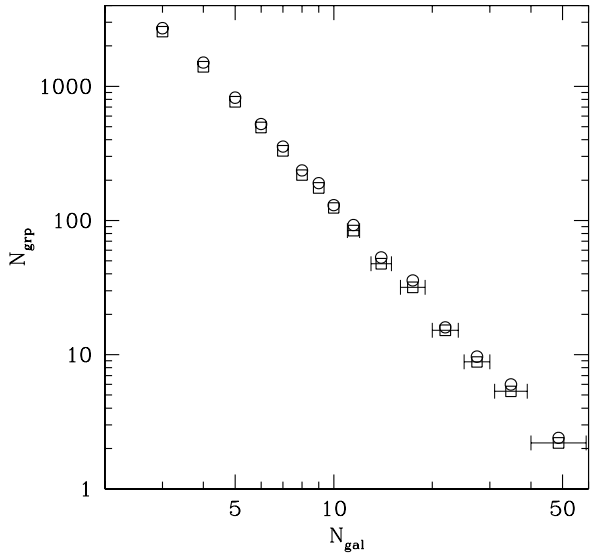} 
\caption{Galaxy group multiplicity function of SDSS Sample II.  Circle
  points show the number of groups as a function of group richness,
  for groups with $M \geq 10^{12} h^{-1} \Msun$ and $50\kms \leq
  \hat{\sigma}_{\rm sat} \leq 1000\kms$.  Square points show $N_{\rm
    grp}(N_{\rm gal})$ for groups in which we additionally require
  that the central galaxy is also the central galaxy of a group in
  Sample III.  Horizontal error bars indicate the width of each bin.}
 \label{SDSSmultfn}
\end{figure}

In Fig.~\ref{SDSSmultfn}, we show the abundance of galaxy groups as
a function of richness in the catalogue, with $M \geq 10^{12}
h^{-1}\Msun$ and $50\kms \leq \hat{\sigma}_{rm sat} \leq 1000\kms$.
The requirement that the brightest group galaxy is also the brightest
group galaxy in Sample III results in slightly fewer groups at all
richnesses. 

The analysis of vdB05 was done with a catalogue of 2502 groups in the
2dFGRS with four members or more; the new Y07 catalogue has nearly twice
as many groups with $N_{\rm gal} \geq 4$ (4571), which is a
significant improvement.  In addition, the typical rms redshift and
magnitude errors of galaxies in the SDSS are $30\kms$ and 0.035 mag
($r$-band), respectively (Strauss \etal 2002), compared to $85\kms$
and 0.15 mag ($B$-band) in the 2dFGRS (Colless \etal 2001).  The
improved statistics of this SDSS DR4 group catalogue allow us to not
only investigate the halo mass dependence of central galaxy velocity
bias, but also the dependence of velocity bias on the properties of
the central galaxies themselves. 

\section{Mock catalogues}
\label{sec:mocks}

The main goal of this paper is to use the distributions of the
parameters $\calR$ and $\calS$ defined above to test our three
hypotheses related to the CGP defined in Section~\ref{sec:intro}. As
discussed above, since the SDSS galaxy group catalogue suffers from
interlopers and incompleteness effects, we require mock group
catalogues constructed from mock galaxy redshift surveys (hereafter
MGRSs) using the same halo-based galaxy group finder as used for the
SDSS.

We construct MGRSs by populating dark matter haloes with galaxies of
different luminosities.  The distribution of dark matter haloes is
obtained from a set of large $N$-body simulations (dark matter only)
for the WMAP3 $\Lambda$CDM cosmology from Macci\`{o} \etal (2007).
The simulations have $512^3$ particles each, have periodic boundary
conditions, and box sizes of $L_{\rm box}=100 h^{-1} \Mpc$ (hereafter
$L_{100}$) and $L_{\rm box}=300 h^{-1} \Mpc$ (hereafter $L_{300}$). We
follow Yang \etal (2004) and replicate the $L_{300}$ box on a $4
\times 4 \times 4$ grid.  The central $2 \times 2 \times 2$ boxes, are
replaced by a stack of $6 \times 6 \times 6$ $L_{100}$ boxes (see
Fig.~11 in Yang \etal 2004).  This stacking geometry circumvents
incompleteness problems in the mock survey due to insufficient mass
resolution of the $L_{300}$ simulations, and allows us to reach the
desired depth of $z_{\rm max}=0.20$ in all directions.

Dark matter haloes are identified using the standard FOF algorithm
with a linking length of $0.2$ times the mean inter-particle
separation.  Unbound haloes and haloes with less than 10 particles are
removed from the sample. The resulting halo mass functions are in
excellent agreement with the analytical halo mass function of Sheth,
Mo \& Tormen (2001).

\subsection{Assigning Luminosities}
\label{sec:lassign}

We populate each halo with galaxies of different luminosities using
the conditional luminosity function (CLF) model described in Cacciato
\etal (2009; hereafter C09). The CLF, $\Phi(L|M)$, specifies the
average number of galaxies of luminosity $L$ in a halo of mass $M$,
and is constrained to accurately match the SDSS $r$-band luminosity
function (Blanton \etal 2003), the clustering strength of SDSS
galaxies as a function of luminosity (Wang \etal 2007), and the
galaxy-galaxy lensing data of Mandelbaum \etal (2006). 
The CLF assumes that the luminosity-dependent abundance, distribution, 
and clustering of galaxies can be described as a function of halo mass.
The CLF of C09 is split in two parts, $\Phi_{\rm cen}(L|M)$ and $\Phi_{\rm
  sat}(L|M)$, which describe the halo occupation statistics of central
and satellite galaxies, respectively.

For each halo we draw the luminosity of its central galaxy from
$\Phi_{\rm cen}(L|M)$, which is parameterized as a log-normal
distribution:
\begin{equation}
\Phi_{\rm cen}(L|M) \rmd L = {1 \over \sqrt{2\pi}\, {\rm ln}(10) \, \sigma_{\rm cen}}
\exp\left[ -\left({\log(L/L_{\rm cen})\over\sqrt{2}\sigma_{\rm cen}}\right)^2\right]
\, {\rmd L \over L}\,.
 \label{CLFcen}
\end{equation}
Here $\sigma_{\rm cen}=0.14$ quantifies the scatter between central
galaxy luminosity and host halo mass.  More \etal (2009) obtained a
similar value from their analysis of satellite kinematics:
$\sigma_{\rm cen}=0.16 \pm 0.04$.  The central galaxy luminosity as a
function of halo mass, $L_{\rm cen}(M)$ is parameterized as a double
power-law, with a slope of $3.3$ in low-mass haloes and $0.26$ in
massive haloes (see C09 for details).

For the satellite galaxies we assume that their halo occupation numbers 
follow a Poisson distribution with mean
\begin{equation}\label{meanNsat}
\langle N_{\rm sat}|M\rangle = \int_{L_{\rm min}}^\infty \Phi_{\rm sat}(L|M) \, \rmd L\,,
\end{equation}
where we adopt a luminosity threshold, $L_{\rm min}$, corresponding to
$M_r - 5\log h = -14$.  The satellite luminosities are drawn from the
satellite CLF $\Phi_{\rm sat}(L|M)$, which is parameterized as a
modified Schechter function:
\begin{equation}
\Phi_{\rm sat}(L|M) = \frac{\phi_{\rm sat}^\ast}{L_{\rm sat}^\ast}
\left({L \over L_{\rm sat}^\ast}\right)^{\alpha_{\rm sat}}
\exp\left[-\left({L\over L_{\rm sat}^\ast}\right)^s\right]\,,
 \label{CLFsat}
\end{equation}
where $\phi_{\rm sat}$ and $\alpha_{\rm sat}$ are functions of halo
mass, the parameter $s=2$, and $L_{\rm sat}^\ast(M)=0.562\,L_{\rm
cen}(M)$ (see C09 for details).  We emphasize that this
particular form for the CLF (eqn.~\ref{CLFcen} and~\ref{CLFsat}) is
not an assumption, but rather is the form that agrees with the CLF
obtained directly from the SDSS group catalogue (see Yang \etal 2008).
The only poorly constrained parameter is $s$, and we test the effect
of its uncertainty in Section~\ref{sec:compCLF}.

If the luminosity of the satellite is brighter than that of its
central, a new luminosity is drawn until it is fainter than that of
the central. Hence, in our mock universe central galaxies are always
BHGs, by construction.

\subsection{Assigning Phase-Space Coordinates}
\label{model}

Having assigned all mock galaxies their luminosities, the next step is
to assign them a position and velocity within their halo.  

We assume that each dark matter halo of mass $M$ has a NFW (Navarro,
Frenk \& White 1997) density distribution, $\rho_{\rm dm}(r|M)$, with
virial radius $r_{\rm vir}(M)$, characteristic scale radius $r_s(M)$,
and concentration parameter $c(M) = r_{\rm vir}/r_s$. We model the
halo concentrations using the $c(M)$ relation of Macci\`{o} \etal
(2007). Assuming haloes to be spherical and isotropic, the local,
one-dimensional velocity dispersion follows from solving the Jeans
equation
\begin{equation}
\label{jeans}
\sigma^2_{\rm dm}(r|M) = {1 \over \rho_{\rm dm}(r|M)} 
\int_{r}^{\infty} \rho_{\rm dm}(r'|M) {\partial \Psi \over \partial
  r}(r'|M) {\rm d}r'
\end{equation}
with $\Psi(r)$ the gravitational potential (Binney \& Tremaine 1987).
Using that $\partial \Psi / \partial r = G M(r)/r^2$ and defining the
virial velocity $V_{\rm vir} = \sqrt{G M / r_{\rm vir}}$ we obtain
\begin{equation}
\label{sig1dm}
\sigma^2_{\rm dm}(r|M) =  V^2_{\rm vir} {c \over f(c)} \, 
\left({r \over r_s}\right) \, \left(1 + {r \over r_s}\right)^2 \, 
{\cal I}(r/r_s)
\end{equation}
with $f(x) = {\rm ln}(1+x) - x/(1+x)$ and
\begin{equation}
\label{cali}
{\cal I}(y) = \int_{y}^{\infty} {f(\tau) \,{\rm d}\tau 
\over \tau^3 (1+\tau)^2}\,.
\end{equation}
The halo-averaged velocity dispersion is given by
\begin{eqnarray}
\label{sigexp}
\langle \sigma_{\rm dm}|M \rangle & \equiv &
{4 \pi \over M}  \int_{0}^{r_{\rm vir}(M)}
\rho_{\rm dm}(r|M) \, \sigma_{\rm dm}(r|M) \, r^2 \, {\rm d}r \nonumber \\
& = &V_{\rm vir} \, \sqrt{c \over f^3(c)} \,
\int_{0}^{c} {y^{3/2} \, {\cal I}^{1/2}(y) \over  
(1+y)} \, {\rm d}y
\end{eqnarray}
(cf. van den Bosch \etal 2004).

\subsubsection{Central Galaxies}

For the central galaxies, we proceed as follows. In the case of
$\calH_0$ and $\calH_2$ mocks, we position the central galaxy at rest
at the centre of the dark matter halo (for the $\calH_2$ mocks, we
then reshuffle the indices of central and brightest satellite, as
detailed below, but we do so only after the construction of the group
catalogue).  In the case of the $\calH_1$ mocks, we follow the
approach of vdB05, to which we refer the reader for details.  Briefly,
we assume that the radial coordinate of the central, $r$, follows a
probability distribution\footnote{The choice for this particular
probability distribution, which corresponds to a Hernquist (1990)
profile, is not motivated by any physical considerations, other than
the fact that it is well behaved, both at $r=0$ and at $r \rightarrow
\infty$.  Our results do not depend significantly on the exact
shape of this probability distribution.}
\begin{equation}
\label{probcen}
P_{\rm cen}(r) {\rm d}r = 2 \left({r_{\rm vir} + a \over r_{\rm vir}} 
\right)^2 {a r \over (r + a)^3} {\rm d} r\,,
\end{equation}
where $a$ is a free parameter, which is related to the velocity bias
$b_{\rm vel}$ as detailed below. In order to parameterize the
characteristic radius $a$ in terms of that of the dark matter halo, we
define the parameter $f_{\rm cen} \equiv a/r_s$.

Central galaxies in a halo of mass $M$ at a halo-centric radius $r$
have an isotropic velocity dispersion
\begin{eqnarray}
\label{veldispcen}
\sigma^2_{\rm cen}(r|M) & = & {1 \over \rho_{\rm cen}(r|M)} 
\int_{r}^{\infty} \rho_{\rm cen}(r'|M) {\partial \Psi \over \partial
  r}(r'|M) {\rm d}r'\nonumber \\
 & = & V^2_{\rm vir} {c \over f(c)} \, 
\left({r \over r_s}\right) \, \left(f_{\rm cen} + {r \over
    r_s}\right)^3 {\cal J}(r/r_s)\,,
\end{eqnarray}
with
\begin{equation}
\label{calj}
{\cal J}(y) = \int_{y}^{\infty} {f(\tau) \,{\rm d}\tau 
\over \tau^3 (f_{\rm cen}+\tau)^3}\,.
\end{equation}
This implies a halo-averaged velocity dispersion of 
\begin{eqnarray}
\label{sigcenav}
\langle \sigma_{\rm cen}|M \rangle & \equiv &
{\int_{0}^{r_{\rm vir}(M)}
\rho_{\rm cen}(r|M) \, \sigma_{\rm cen}(r|M) \, r^2 \, {\rm d}r 
\over \int_{0}^{r_{\rm vir}(M)} \rho_{\rm cen}(r|M) \, r^2 \, {\rm d}r} \nonumber \\
& = &V_{\rm vir} \, \sqrt{4 c \over f(c)} \, f_{\rm cen} \,
\int_{0}^{c} {y^{3/2} \, {\cal J}^{1/2}(y) \over  
(f_{\rm cen}+y)^{3/2}} \, {\rm d}y\,,
\end{eqnarray}
which allows us to define the velocity bias of central galaxies as
\begin{equation}
b_{\rm vel}  \equiv {\langle\sigma_{\rm cen}|M \rangle \over  
\langle\sigma_{\rm dm}|M  \rangle} = {\langle\sigma_{\rm cen}|M \rangle
\over \langle\sigma_{\rm sat}|M \rangle}\,.
\end{equation} 
In addition to the velocity bias, we define the spatial bias as
\begin{equation}
b_{\rm  rad} \equiv {\langle r_{\rm cen}|M \rangle \over
\langle r_{\rm  dm}|M \rangle}  = {\langle r_{\rm  cen}|M \rangle \over 
\langle r_{\rm  sat}|M \rangle}\,,
\end{equation}
where the expectation value for the radius follows from
\begin{equation}
\label{rexpect}
\langle r|M \rangle = {\int_{0}^{r_{\rm vir}} \rho(r) r^3 {\rm d}r 
\over \int_{0}^{r_{\rm vir}} \rho(r) r^2 {\rm d}r}
\end{equation}
For comparison, an NFW density distribution has $\langle r|M \rangle =
0.41 r_{\rm vir}(M)$ for $c=10$, and $0.47 r_{\rm vir}(M)$ for $c=5$.
\begin{figure}
\includegraphics[width=\hsize]{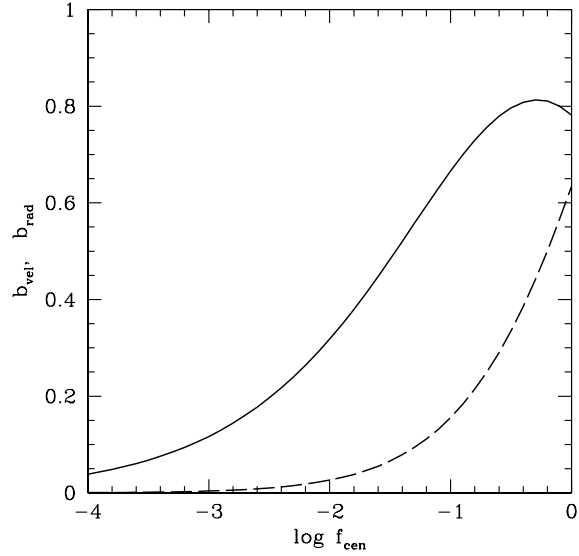} 
\caption{The velocity bias (solid curve) and spatial bias (dashed
  curve) of central galaxies as a function of the parameter $f_{\rm
    cen}$, which expresses the characteristic scale of the radial
  distribution of central galaxies in terms of the characteristic
  scale of the NFW density distribution.  The results shown correspond
  to a dark matter halo with a concentration $c=10$.}
 \label{bvelfcen}
\end{figure}

With this model, a particular value of the parameter $f_{\rm cen}$
implies a particular amount of velocity and spatial bias.  The
relations $b_{\rm vel}(f_{\rm cen})$ and $b_{\rm rad}(f_{\rm cen})$
are shown in Figure~\ref{bvelfcen}.  These relations only depend
weakly on halo concentration (see vdB05).  In the limit $f_{\rm cen}
\rightarrow 0$, the probability distribution $P_{\rm cen}(r)$ becomes
a delta function, implying that the central galaxy is sitting at rest
at the centre of the dark matter halo (i.e., the null-hypothesis
$\calH_0$ of the CGP).  Larger values of $f_{\rm cen}$ result in
larger amounts of velocity and spatial bias. Note that $b_{\rm vel}$
is always larger than $b_{\rm rad}$, indicating that the signature of
an off-centered central galaxy (i.e., hypothesis $\calH_1$) is more
pronounced, and thus easier to detect, in velocity space than in
configuration space.

\subsubsection{Satellite Galaxies}

Throughout this paper, we assume that the $N_{\rm sat}$ satellite
galaxies in a halo of mass $M$ follow a number density distribution
$n_{\rm sat}(r|M) = (N_{\rm sat} / M) \rho_{\rm dm}(r|M)$, so that there is
no spatial bias between satellite galaxies and dark matter particles.
If we further assume that the satellites are in isotropic equilibrium,
it also follows that there is no velocity bias between the satellites
and the dark matter, neither globally [i.e., $\langle \sigma_{\rm sat}|M
\rangle = \langle \sigma_{\rm dm}|M \rangle$] nor locally [i.e.
$\sigma_{\rm sat}(r|M) = \sigma_{\rm dm}(r|M)$].

For simplicity, we assume that the satellites follow a
spherically symmetric spatial distribution with a velocity
distribution that is locally isotropic.  Although this is a clear
oversimplification, 
as some galaxy groups and clusters have anisotropic or aspherical 
distributions (e.g., Bailin et al. 2008; Wang et al. 2008),
we do not believe that it strongly impacts our
results.  For example, the global velocity dispersion (i.e., obtained
from all satellites) of an anisotropic system will be nearly identical
to that of an isotropic system with the same gravitational potential,
since both are governed by the virial equation.  In other words,
anisotropy changes the {\it local} line-of-sight velocity distribution
(LOSVD), but leaves the second moment of the {\it global} LOSVD
largely unchanged (see, e.g., van den Bosch \etal 2004).  Therefore,
since the $\calR$ parameter (eqn.~\ref{Rparam}) only depends on the
first and second moments of the global LOSVD, our results are robust
to the assumptions of isotropy and sphericity.
The cumulative $\calR$ distribution is weakly dependent on higher moments 
of the LOSVD, but again, the effect of anisotropy is very small.

Although there is evidence to support our assumption that satellite
galaxies have number density distributions that are well fit by a NFW
profile (e.g., Lin \etal 2004), the corresponding concentration
parameters seem to be significantly smaller than the values expected
for their dark matter haloes, by about a factor of 2 to 3 (e.g.,
Hansen \etal 2005; Yang \etal 2005b).  Since we assume that
satellites are unbiased with respect to the dark matter, our value for
$b_{\rm vel}$ will be incorrect by a factor of $\langle\sigma_{\rm
  sat}\rangle / \langle\sigma_{\rm dm}\rangle$.  Using the model
described in Eqs~(\ref{sigexp}) and~(\ref{sigcenav}), we estimate that
this may result in an underestimating of the velocity bias by no more
than 5\%.  Given that this is a small effect compared to the other
uncertainties involved in our analyses, we do not attempt to
correct for it.

\subsection{Mock Group Catalogues}

Once the dark matter haloes are populated with galaxies, we construct
a MGRS following the procedure described in Li \etal (2007). We place
a virtual observer at the centre of the stack of simulation boxes,
assign each galaxy a ($\alpha$, $\delta$)-coordinate, and remove the
ones that are outside the mocked SDSS survey region.  For each model
galaxy in the survey region, we compute its redshift (which includes
the cosmological redshift due to the universal expansion, the peculiar
velocity, and a $35\kms$ Gaussian line-of-sight velocity dispersion to
mimic the redshift errors in the data), and its $r$-band apparent
magnitude (based on the $r$-band luminosity of the galaxy).

We eliminate galaxies that are fainter than the SDSS apparent
magnitude limit, and incorporate the position-dependent incompleteness
by randomly eliminating galaxies according to the completeness factors
obtained from the survey masks provided by the NYU-VAGC (Blanton \etal
2005).  Finally, we construct group catalogues from the MGRSs, using
the same halo-based group finder as used for the real SDSS DR4. 

Using the method outlined above, we construct the following set of
mock group catalogues. The first is a set of ten mocks that only differ
in the value of the velocity bias $b_{\rm vel} = 0.0, 0.1, 0.2, 0.3,
0.4, 0.5, 0.6, 0.7, 0.8$, and $1.0$. These constitute our $\calH_1$
mocks. Note that the mock with $b_{\rm vel}=0$ satisfies the CGP, and
therefore corresponds to the null-hypothesis $\calH_0$, while the
other mocks correspond to $\calH_1$ (i.e., BHG is central galaxy, but
is not at rest at centre of dark matter potential well). The second
set of mock group catalogues is constructed starting from the $b_{\rm
  vel}=0$ mock group catalogue, in which we switch the luminosities of
the central and its brightest satellite for a random fraction $\fFU$
of all groups.  We construct 6 mock group catalogues for $\fFU = 0.1,
0.2, 0.3, 0.4, 0.5$, and $0.6$, respectively. These constitute our
$\calH_2$ mocks.

In order to facilitate the comparison with the SDSS galaxy group
catalogue, in what follows we discard all mock groups with an assigned
group mass less than $10^{12} h^{-1} \Msun$, and with a satellite
velocity dispersion $\hat{\sigma}_{\rm sat} < 50\kms$ or
$\hat{\sigma}_{\rm sat} > 1000\kms$. For the mock groups that have
not been discarded, we compute the parameters $\calR$ and $\calS$.  In
the next section, we compare the resulting distributions
$P(<\vert\calR\vert)$ and $P(<\vert\calS\vert)$ with those obtained
from the SDSS group catalogue in order to test our three hypotheses
$\calH_0$, $\calH_1$, and $\calH_2$.
\begin{figure}
 \includegraphics[width=\hsize]{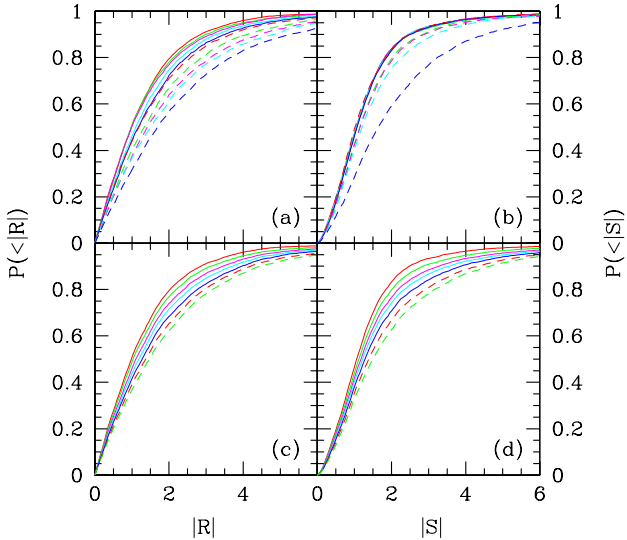} 
 \caption{The cumulative distributions of $\vert\calR\vert$ (left
   panels) and $\vert\calS\vert$ (right panels) for the groups in mock
   group catalogues with four members or more.  Upper panels show the
   distributions for mock catalogues with different amounts of $b_{\rm
     vel}$: the solid curves are for $b_{\rm vel}=0$, 0.1, 0.2, 0.3,
   and 0.4 (red, green, magenta, cyan, and blue curves, respectively)
   and the dashed curves are for $b_{\rm vel}=0.5$, 0.6, 0.7, 0.8, and
   1 (red, green, magenta, cyan, and blue curves, respectively).
   Lower panels show the distributions for mock catalogues with
   different fractions $\fFU=0$, 0.1, 0.2, 0.3, 0.4, 0.5, and 0.6
   (red, green, magenta, cyan, blue, dashed red, and dashed green
   curves, respectively).}
 \label{mocksbdep}
\end{figure}

Finally, we note that the halo centres and central and satellite
galaxies are defined only in the mock group catalogues, not in the
SDSS group catalogue.  The $\mathcal{R}$ and $\mathcal{S}$ parameters
used in the following analysis are defined with respect to the BHGs in
the mock and observed galaxy groups.  This way, we are able to
statistically analyze groups of a wide range in mass, including poor
groups for which a centre or central galaxy might not be well-defined.

\section{Results}
\label{sec:res}

Fig.~\ref{mocksbdep} shows the cumulative distributions of
$\vert\calR\vert$ (left panels) and $\vert\calS\vert$ (right panels)
obtained from the $\calH_1$ mocks (upper panels) and the $\calH_2$
mocks (lower panels) for groups with four or more members (i.e.,
$N_{\rm sat} \geq 3$). As expected, $P(\vert\calR\vert)$ and
$P(\vert\calS\vert)$ become broader for larger values of $b_{\rm vel}$
(upper panels) and $\fFU$ (lower panels). A few trends are
apparent. Firstly, note that the $\vert\calS\vert$-distributions
are all bunched together for $b_{\rm vel} \leq 0.5$. Only for $b_{\rm
  vel} > 0.5$ are the cumulative $\vert\calS\vert$-distributions
notably different.  This shows that it is difficult to constrain
$b_{\rm vel}$ using the angular positions of galaxies, on which the
$\calS$-statistic is based, vis-\`{a}-vis using the velocity-statistic
$\calR$. Physically this is a reflection of the fact that dark matter
haloes have steep potential wells, so that large velocities are
required even for relatively modest excursions from the
centre. Secondly, and most importantly, comparing the upper and lower
panels, it is clear that a non-zero value of $\fFU$ has a different
impact on $P(<\vert\calR\vert)$ and $P(<\vert\calS\vert)$ than a
non-zero value of $b_{\rm vel}$. This is good news, as it implies that
we will be able to discriminate between our three hypotheses.
\begin{table}
 \begin{center}
  \begin{tabular}[h!]{ l | c c c c }
   Halo Mass Bins & $\mathrm{log}\,M_\mathrm{min}$ & ${\overline{\log M}}$ & $\mathrm{log}\,M_\mathrm{max}$ & $N_\mathrm{group}$ \\
   \hline
   low mass & 12 & 12.86 & 13.3 & 1434 \\
   intermediate mass & 13.3 & 13.53 & 13.75 & 1540 \\
   high mass & 13.75 & 14.09 & 15.2 & 1555 \\
   \hline
  \end{tabular}
 \end{center}
 \caption{Halo mass bins ($\log(M/h^{-1}\Msun)$) in the SDSS group catalogue, 
   for groups with four members or more and 
   $50\,\kms \leq \hat{\sigma}_{\rm sat} \leq 1000\,\kms$. 
   The log mass ranges, mean masses, and number of groups are given.}
 \label{Mgrpdist}
\end{table}

In what follows we will compare the cumulative $\vert\calR\vert$ and
$\vert\calS\vert$ distributions obtained from the SDSS with those
obtained from our mock catalogues. Since we have a relatively large
number of groups in our SDSS sample, we can perform this test for a
(small) number of bins in group mass, thus giving some leverage on a
possible halo mass dependence.  At the massive end of the group mass
distribution, the group catalogue is roughly complete, and the mass
distribution closely follows the halo mass function. The constraint of
four or more group members, however, cuts off the distribution at the
low mass end, leaving virtually no groups with $M < 10^{12} h^{-1}
\Msun$. We split the SDSS group catalogue in three mass bins (listed
in Table~\ref{Mgrpdist}) that contain a similar number ($\sim 1500$)
of groups, and the corresponding mass cuts occur at
$\log(M/h^{-1}\Msun) = 13.3$ and $13.75$.

The results presented in Sections~\ref{sec:testH1} and
\ref{sec:testH2} are based on groups with at least four members (i.e.,
with $N_{\rm sat} \geq 3$).  We have repeated these analyses using
samples of groups with three, five, six, and ten or more members.  We
have also tested lower and higher halo mass thresholds.  In each case
we obtain constraints on $b_{\rm vel}$ and $\fFU$ that are consistent
with each other at the $1\sigma$ level.  Hence, our results do not
depend significantly on the multiplicity and halo mass thresholds
used.

\subsection{Testing Hypothesis $\calH_1$}
\label{sec:testH1}

\begin{figure*}
\includegraphics[width=0.497\hsize]{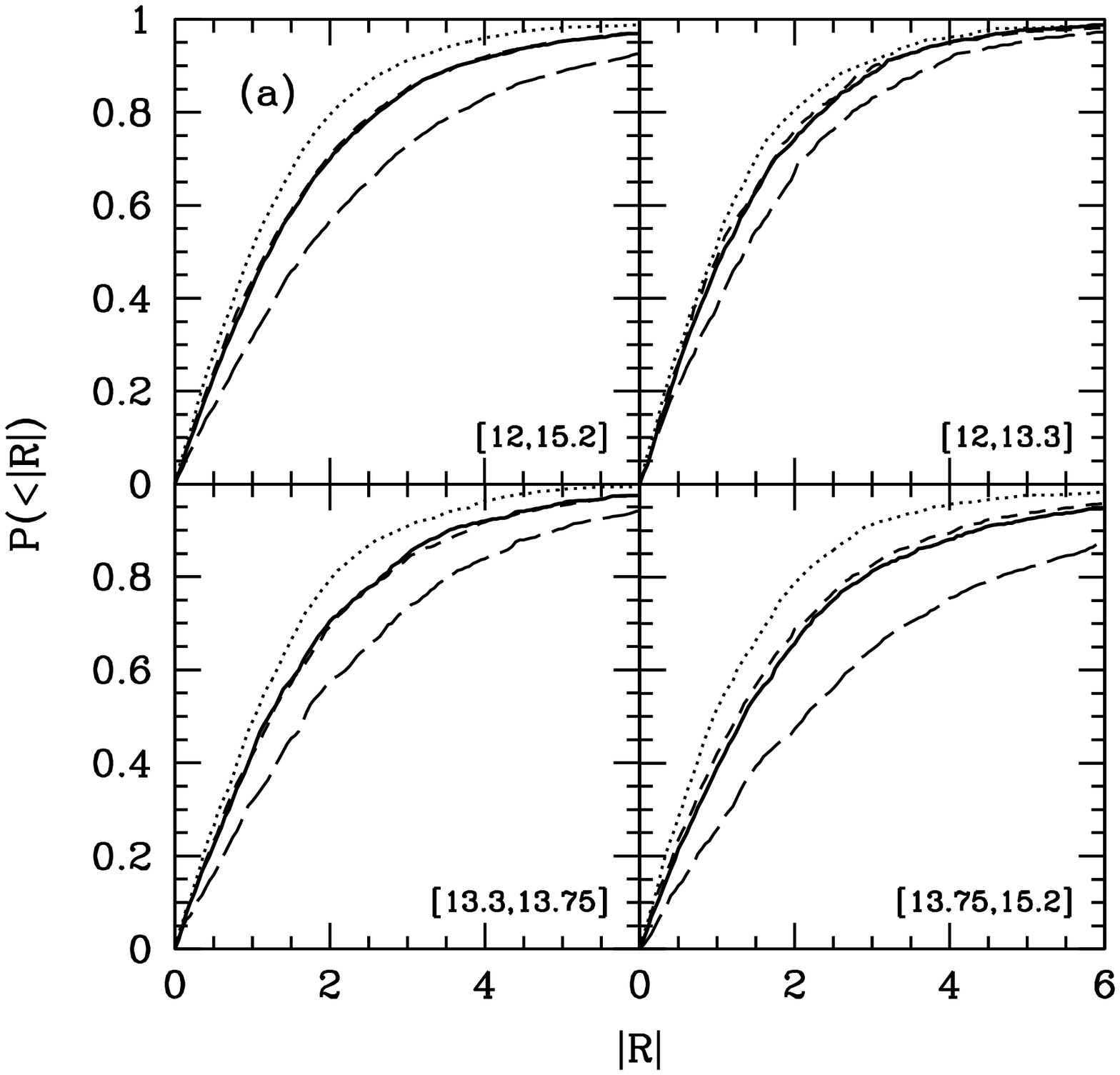} 
\includegraphics[width=0.497\hsize]{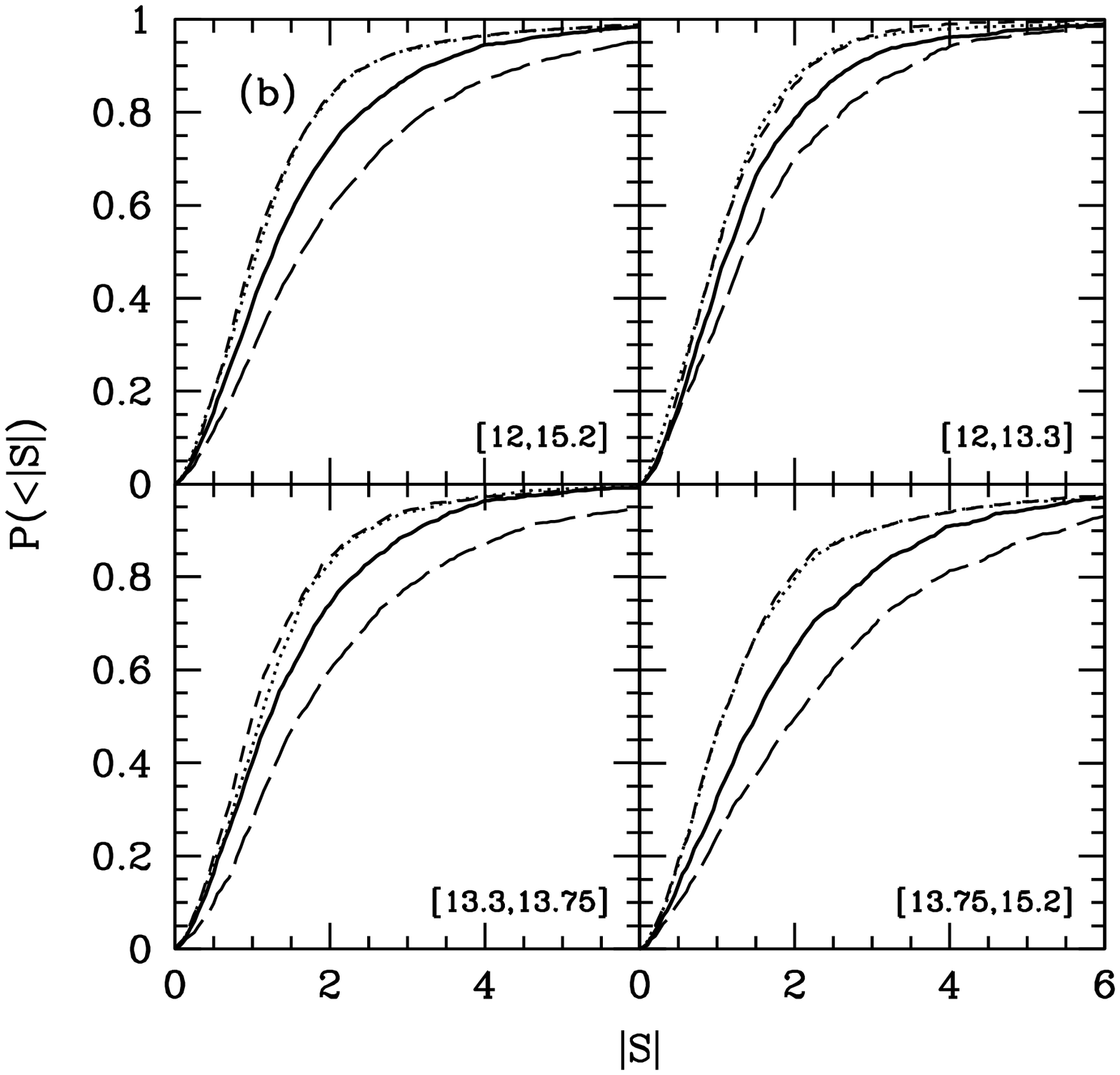} 
\caption{The cumulative distribution of $|\calR|$ (left figure) and
   $|\calS|$ (right figure) obtained from the SDSS group catalog
   (solid black curve) compared with those obtained from three of our
   mocks, with $b_{\rm vel}=0$ (red dotted curve), 0.5 (blue
   short-dashed curve), and 1 (green long-dashed curve), for groups
   with four members or more.  Results are shown for four log halo
   mass intervals, as indicated in square brackets in each panel.}
\label{MhdepPofR}
\end{figure*}

In order to test hypothesis $\calH_1$, we compare the SDSS group
catalogue to mock group catalogues with different values of $b_{\rm
vel}$ while setting $\fFU=0$.  Figure~\ref{MhdepPofR}a shows the
cumulative $\vert\calR\vert$-distribution for different bins of halo
mass obtained from the SDSS group catalogue (solid lines) and from
three mock group catalogues corresponding to different values of
$b_{\rm vel}$ (0, 0.5, and 1).  In all cases only groups with four
members of more are considered. The numbers in square brackets in each
panel indicate the range of $\log(M/h^{-1}\Msun)$ considered.  Note
that at fixed $\calR$, the corresponding $P(<\vert\calR\vert)$ becomes
smaller when using a bin with larger group masses. This does not
necessarily imply a trend of $b_{\rm vel}$ with halo mass, though. It
may also be due to a mass dependence of the fraction of interlopers,
or a mass dependence of the completeness of group members. Indeed, the
mock group samples for a fixed value of $b_{\rm vel}$ reveal the same
trend, indicating that it is most likely an artifact introduced by the
group finder. For each halo mass bin, the mock catalogue with $b_{\rm
vel}=0$ (i.e., the one that fulfills the null-hypothesis, $\calH_0$,
of the CGP) predicts a $\vert\calR\vert$-distribution that is much
narrower than that of the SDSS, while the mock catalogue with $b_{\rm
vel}=1$ yields a distribution that is too broad.  Interestingly, the
intermediate case, with $b_{\rm vel}=0.5$, results in
$P(<\vert\calR\vert)$ that are very similar to those of the SDSS, for
each mass bin.  This rules out the null hypothesis $\calH_0$, and
seems to suggest that instead central galaxies have a relatively large
velocity bias with $\langle \sigma_{\rm cen}|M \rangle \simeq 0.5
\langle \sigma_{\rm sat}|M \rangle$ with little dependence on halo
mass $M$.

However, Fig.~\ref{MhdepPofR}b, which is similar to
Fig.~\ref{MhdepPofR}a except that it shows the cumulative
$\vert\calS\vert$-distributions, paints a different picture.  Both the
$b_{\rm vel} = 0.0$ and $b_{\rm vel} = 0.5$ mocks predict
$\vert\calS\vert$-distributions that are clearly too narrow, for each
mass bin. Rather, the SDSS $\vert\calS\vert$-distribution seems to
require $0.5 < b_{\rm vel} < 1.0$. Hence, it appears that no single
value of $b_{\rm vel}$ can {\it simultaneously} match
$P(<\vert\calR\vert)$ and $P(<\vert\calS\vert)$ obtained from the
SDSS, ruling against our hypothesis $\calH_1$.  In other
words, although the BHGs have non-zero velocities, the velocity spread
is too small to explain their displacements from the halo centres.
\begin{figure}
\includegraphics[width=\hsize]{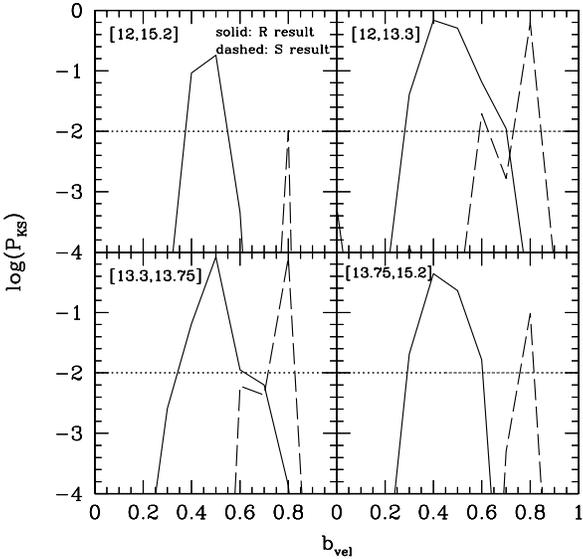} 
\caption{The KS-probability that the cumulative $\calR$ distribution
  (solid lines) and $\calS$ distribution (dashed lines) obtained from
  the SDSS groups is consistent with that obtained from our mocks, as
  a function of $b_{\rm vel}$.  Results are shown for four log halo
  mass intervals, indicated in square brackets in each panel.  The
  horizontal dotted line in each panel indicates $P_{\rm KS}=0.01$:
  based on estimates of the scatter due to cosmic variance, we
  consider two distributions to be statistically equivalent when
  $P_{\rm KS}>0.01$.  }
 \label{Mhdepbvel}
\end{figure}

In order to make this more quantitative, we use the Kolmogorov-Smirnov
(hereafter KS) test to compute the probabilities $P_{\rm KS}$ that
$P(\vert\calR\vert)$ and $P(\vert\calS\vert)$ of the SDSS and a
particular mock are drawn from the same distribution.  The results are
shown in Fig.~\ref{Mhdepbvel}, which plots $\log(P_{\rm KS})$ as
function $b_{\rm vel}$ for each of the four mass bins considered.
Using the median and the 16 and 84 percentiles of $P_{\rm KS}(b_{\rm
  vel})$ for the $\vert\calR\vert$ distributions (solid lines) we
obtain $b_{\rm vel}=0.47_{-0.07}^{+0.06}$ for the entire mass range,
and $b_{\rm vel}=0.44_{-0.07}^{+0.09}$, $b_{\rm vel}=0.50\pm0.05$, and
$b_{\rm vel}=0.43_{-0.06}^{+0.08}$ for the low-mass,
intermediate-mass, and high-mass intervals, respectively. For the
$\vert\calS\vert$ distributions (dashed lines) this yields $b_{\rm
  vel}=0.82_{-0.06}^{+0.08}$ for the low-mass interval, and $b_{\rm
  vel}=0.83_{-0.06}^{+0.08}$ for the intermediate- and high-mass
intervals.  Therefore, the amounts of velocity bias consistent with
the $\calR$ and $\calS$ distributions of SDSS groups are significantly
different, with their best-fit values of $b_{\rm vel}$ differing from
each other by more than 4$\sigma$.  We conclude that hypothesis
$\calH_1$ is ruled out, since it cannot simultaneously explain the
$\vert\calR\vert$ and $\vert\calS\vert$ distributions obtained from
the SDSS group catalogue.

\subsection{Testing Hypothesis $\calH_2$}
\label{sec:testH2}

\begin{figure}
\includegraphics[width=\hsize]{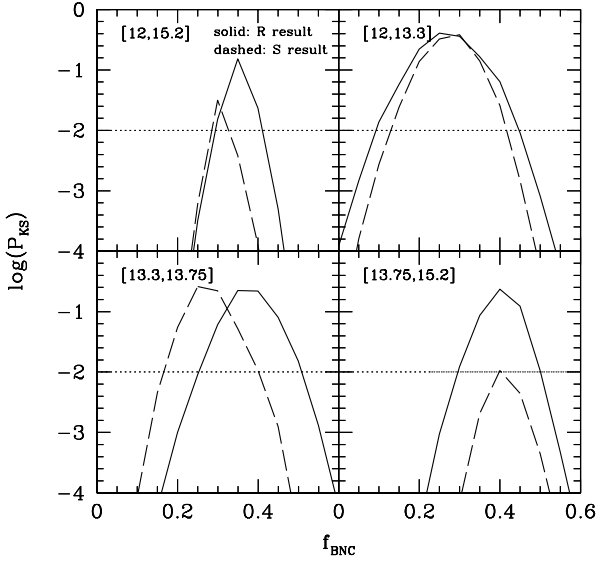} 
\caption{The KS-probability that the cumulative $\calR$ distribution
  (left figure) and $\calS$ distribution (right figure) obtained from
  the SDSS groups with four or more members is consistent with that
  obtained from MGRSs, as a function of $\fFU$, the fraction of groups
  in which the most luminous satellite is brighter than the central
  galaxy (see text for details).  The KS probabilities here are the means
  of $\mathrm{log}(P_{\rm KS})$ for 100
  realizations with different random seeds.}
 \label{MhdepfracFUadded} 
\end{figure}

Having ruled out both $\calH_0$ and $\calH_1$, we now turn our
attention to hypothesis $\calH_2$ and investigate whether there is a
value of $\fFU$ for which the mock group catalogue yields
$\vert\calR\vert$- and $\vert\calS\vert$-distributions that are
consistent with the SDSS data.  To that extent we compare the SDSS
group catalogue to mock catalogues with $b_{\rm vel}=0$, but with
different values of $\fFU$.  We proceed in the same way as for
$\calH_1$ in the previous section: for each mock, which corresponds to
a different value of $\fFU$, we compute $P(<\vert\calR\vert)$ and
$P(<\vert\calS\vert)$, which we compare to the corresponding
distributions obtained from the SDSS, resulting in a value for $P_{\rm
  KS}$, the KS-probability that the mock and SDSS distributions are
consistent with each other.

The results are shown in Fig.~\ref{MhdepfracFUadded}, where the solid
and dashed lines once again correspond to the $\calR$ and $\calS$
distributions, respectively.  These KS probabilities are the
means of $\mathrm{log}(P_{\rm KS})$ for 100 realizations with
different random seeds.  The solid lines show that the $\calR$
distribution of SDSS groups indicates a relatively large fraction of
groups in which the central galaxy is not the BHG, with $\fFU$
increasing from $\sim 0.25$ in low mass haloes ($10^{12} h^{-1} \Msun
\leq M \lta 2 \times 10^{13} h^{-1}\Msun$) to $\sim 0.4$ in massive
haloes ($M \ga 5 \times 10^{13.75} h^{-1} \Msun$). Interestingly, the
KS probabilities obtained from the $\calS$ distributions shown by the
dashed lines yield best-fit values of $\fFU$ that are very
similar. Hence, we conclude that contrary to hypothesis $\calH_1$,
hypothesis $\calH_2$ can simultaneously match the
$P(<\vert\calR\vert)$ and $P(<\vert\calS\vert)$ obtained from the SDSS
group catalogue.

\begin{figure}
\includegraphics[width=\hsize]{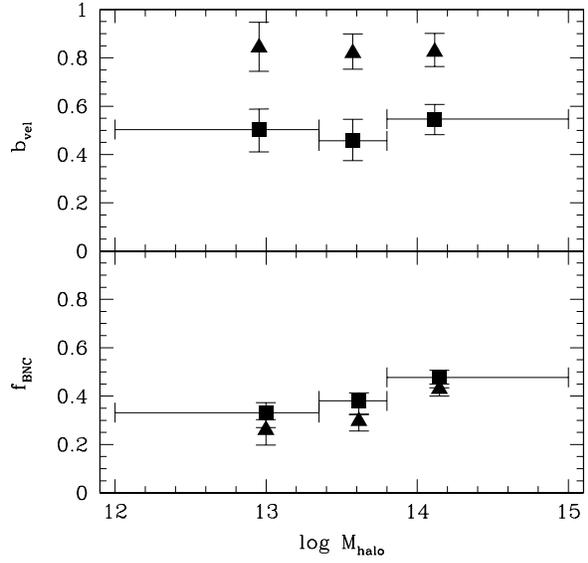} 
\caption{Halo mass dependence of $b_{\rm vel}$ (upper panel) and
   $\fFU$ (lower panel), inferred from the
   $\vert\calR\vert$-distributions (squares) and
   $\vert\calS\vert$-distributions (triangles) obtained from the
   analyses of the SDSS group catalogue (see text for details).  For
   the results in the upper panel, the mock catalogues had different
   values of $b_{\rm vel}$ and $\fFU=0$; for the lower panel, the
   mocks had different values of $\fFU$ and $b_{\rm vel}=0$.  Points
   show best-fit values of the parameters estimated from the KS
   probabilities, vertical error bars show 1-$\sigma$ uncertainties
   from the 16 and 84 percentiles of the $P_{\rm KS}$
   distributions, and horizontal error bars indicate the widths of
   the mass bins.  Hypothesis $\calH_1$ is clearly ruled out from the
   fact that the $\calR$-data data implies a velocity bias, $b_{\rm
   vel}$, that is inconsistent with the value inferred from the
   $\calS$-data.}
 \label{bvelfFUvsM}
\end{figure}

Fig.~\ref{bvelfFUvsM} summarizes our results. It shows the best-fit
values of $b_{\rm vel}$ (upper panel) and $\fFU$ (lower panel) as
functions of halo mass, as inferred from the
$\vert\calR\vert$-distributions (squares) and
$\vert\calS\vert$-distributions (triangles) obtained from the SDSS
galaxy group catalogue.  The figure clearly shows that $\calH_2$
(i.e., a non-zero $\fFU$) can simultaneously explain both the
velocities and the positions of BHGs with respect to their satellites,
while $\calH_1$ (i.e., a non-zero $b_{\rm vel}$) is ruled out because
it cannot.

Considering the $\vert\calR\vert$- and $\vert\calS\vert$-distributions
simultaneously, we infer best-fit values for the fraction of haloes in
which centrals are not BHGs of $\fFU=26_{-6}^{+4}\%$ for $12 \leq
\log(M/h^{-1}\Msun) < 13.3$, increasing to $\fFU=30_{-4}^{+3}\%$ for
haloes with $13.3 \leq \log(M/h^{-1}\Msun) < 13.75$ and
$\fFU=43_{-3}^{+2}\%$ in the most massive haloes with
$\log(M/h^{-1}\Msun) \geq 13.75$.  As in
Section~\ref{sec:testH2}, the best-fit values and uncertainties are
determined from the median and the 16 and 84 percentiles of the
$P_{\rm KS}(\fFU)$ distributions.

Before proceeding with a discussion regarding the implications of
these findings, we briefly address the possibility that both $\calH_1$
and $\calH_2$ are true; BHGs are not central galaxies in a non-zero
fraction $\fFU$ of all haloes, {\it and} central galaxies have a
non-zero value for their velocity bias $b_{\rm vel}$.  Using mock
group catalogues with non-zero values for both $\fFU$ and $b_{\rm
  vel}$ we estimate that the data can accommodate a small amount of
velocity bias $b_{\rm vel} \lta 0.2$. We emphasize, though, that the
data do not {\it require} a non-zero $b_{\rm vel}$.

\subsection{The Impact of Substructure}
\label{sec:substructure}

In the tests described above, we have always assumed that satellite
galaxies follow a smooth, spherically symmetric, number density
distribution, and we assigned the satellites peculiar velocities
within its host halo under the assumption that the corresponding
potential is smooth. This ignores the fact that dark matter haloes are
believed to have a wealth of substructure (see Giocoli et al. 2010,
and references therein; on substructure in galaxy clusters, see Richard et al. 2010). 
Satellite galaxies are believed to
be associated with these substructures. Our treatment of the spatial
and kinematic properties of satellite galaxies is only consistent with
this concept of substructure if: (i) each subhalo hosts at most one
satellite, and (ii) the positions and velocities of subhaloes are not
correlated with those of other subhaloes. In reality, neither of these
criteria is likely to be met. Massive subhaloes are likely to host
multiple satellites, and the large scale filamentary structure is
believed to introduce (weakly) correlated directions of infall for
subhaloes (e.g., Vitvitska \etal 2002; Aubert, Pichon \& Colombi
2004; White et al. 2010). Both of these effects result in correlations among the
positions and/or velocities of different satellite galaxies within the
same host halo, which is likely to have an impact on the $\calR$ and
$\calS$ statistics.

\begin{figure}
 \includegraphics[width=\hsize]{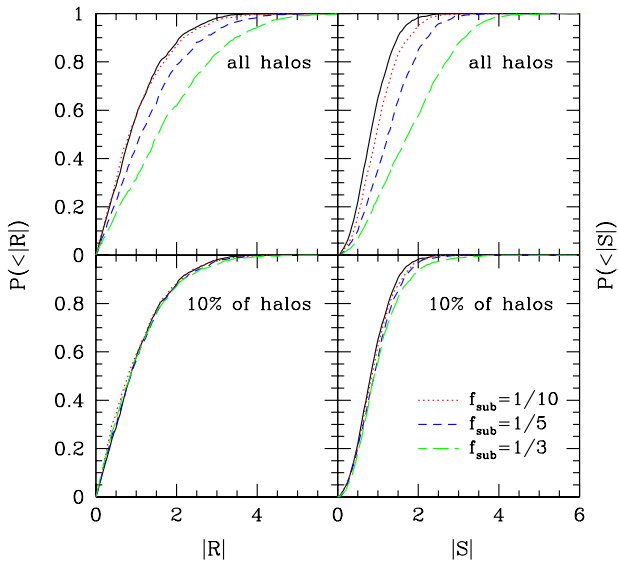} 
 \caption{Cumulative distributions of $\vert\calR\vert$ (left panels) and 
   $\vert\calS\vert$ (right panels) of 1000 mock haloes with 
   $M = 3 \times 10^{14} h^{-1}\Msun$ populated with galaxies.
   Black solid curves indicate the distributions for haloes with no substructure, 
   while the long-dashed green curves, short-dashed blue curves, and dotted red curves 
   indicate the distributions for $f_{\rm sub} = 1/3$, $1/5$, and $1/10$, respectively.
   Results are shown for the case in which all haloes have substructure (upper panels) 
   and $10\%$ of the haloes have substructure (lower panels).}
 \label{subtest}
\end{figure}

In order to have a crude estimate of the impact of substructure, we
proceed as follows. We start by populating 1000 (virtual) dark matter
haloes, all with $M = 3 \times 10^{14} h^{-1}\Msun$, with galaxies
using the same CLF model as in Section~\ref{sec:mocks}.  Similar to the
$\calH_0$ mocks, we assure that the central galaxy is always the
brightest galaxy in the halo, and we position it at rest at the centre
of the halo.  We then compute $\calS$ and $\calR$ for each of these
haloes.  The black, solid lines in Fig.~\ref{subtest} indicate the corresponding
$P(<\vert\calR\vert)$ and $P(<\vert\calS\vert)$. Next we repeat this
exercise, but now, in each halo, a fraction $f_{\rm sub}$ of all
satellites is `clumped' together in a single substructure. We model
this substructure as a halo of mass $m = f_{\rm sub} M$ (i.e., we
assign the galaxies in this subhalo phase-space coordinates in exactly
the same way as we would if the halo was a `host' halo of the same
mass).  The phase-space coordinates of the centre of the subhalo are
drawn in the same way as the phase-space coordinates of the satellites
that are not in a substructure.  The long-dashed green curves, short-dashed
blue curves, and dotted red curves in Fig.~\ref{subtest} show the
$P(<\vert\calR\vert)$ and $P(<\vert\calS\vert)$ thus obtained for
$f_{\rm sub} = 1/3$, $1/5$, and $1/10$, respectively. A comparison
with the no-substructure case (black, solid line) shows that
substructure significantly affects the $\calR$ and $\calS$
distributions, and hence our conclusions, if {\it all} dark matter
haloes have a most massive substructure whose mass is $m_{\rm sub}
\gta 0.1 M$.  In particular, if many haloes contain substantial
substructure, then an alternative explanation for the fact that our
fiducial model (which follows the CGP) does not match the data may be
that the phase-space coordinates of satellite galaxies within the same
dark matter host halo are correlated.  In that case the fractions
$\fFU$ obtained in Section~\ref{sec:testH2} will be significantly
overestimated.

However, using the subhalo mass functions of Giocoli \etal (2010), we
estimate that only $\sim 8\%$ of host haloes with $M = 3 \times
10^{14} h^{-1}\Msun$ have a most massive subhalo with $f_{\rm sub} =
m/M \geq 0.1$, while only $\sim 0.7$ percent have a most massive
subhalo with $m/M \geq 0.3$. Based on these estimates, and on the
tests described above, we argue that ignoring substructure when
populating haloes with (satellite) galaxies does {\it not} have a
significant impact on the $\calR$ and $\calS$ statistics, though more
sophisticated tests are required to confirm this.  Therefore, we
conservatively argue that the $\fFU$ fractions obtained in this paper
have to be regarded as upper limits.

\section{Discussion}
\label{sec:disc}

The results presented in the previous section suggest that in as
  much as 25 to 40 percent of all haloes, the brightest galaxy is a
satellite rather than a central galaxy.  In order to put these numbers
in perspective, we compare them to predictions from halo occupation
statistics (Section~\ref{sec:compCLF}) and from two semi-analytical
models of galaxy formation (Section~\ref{sec:compSAM}). We also
investigate the impact of non-zero $\fFU(M)$ on the halo masses
inferred using satellite kinematics (Section~\ref{sec:satkin}).

\subsection{Comparison with Halo Occupation Statistics}
\label{sec:compCLF}

\begin{figure}
\includegraphics[width=\hsize]{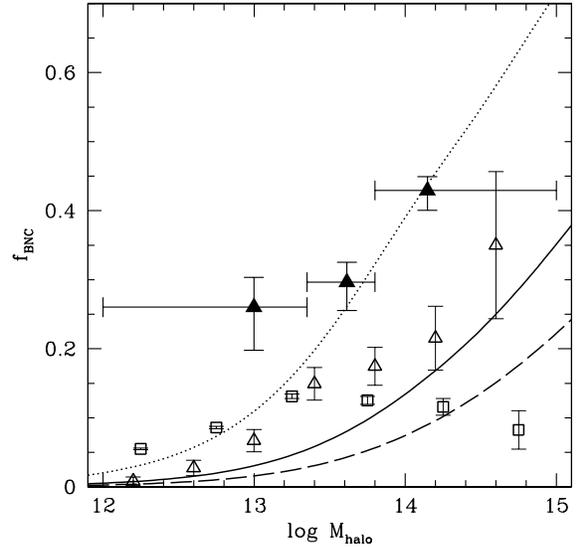} 
\caption{Probability that the most luminous satellite galaxy is
   brighter than the central galaxy in a halo, as a function of halo
   mass (Eqn.~\ref{probLsat1M}).  Result is shown for three slopes in
   the satellite CLF (Eqn.~\ref{CLFsat}): $s=1$ (dotted curve), the
   fiducial $s=2$ (solid curve), and $s=3$ (dashed curve).  For
   comparison, the $\fFU(M_{\rm halo})$ result from
   Figure~\ref{bvelfFUvsM} (solid triangle points) is also shown.  The
   predictions from the \textsc{morgana} semi-analytic model (open
   triangles) and Croton et al. (2006) semi-analytic model (open squares) 
   are also shown, with Poisson errors.  }
 \label{probLsat1Mgrp}
\end{figure}

Using the SDSS $r$-band luminosity function of Blanton \etal (2003),
the projected two-point correlation functions of Wang \etal (2007),
and the galaxy-galaxy lensing data of Mandelbaum \etal (2006), C09
constrained the halo occupation statistics as parameterized by the
conditional luminosity function (CLF)
\begin{equation}
\Phi(L|M) = \Phi_{\rm cen}(L|M) + \Phi_{\rm sat}(L|M)\,,
\end{equation}
which is found to be in good agreement with direct constraints from
the Y07 group catalogue (Yang, Mo \& van den Bosch 2008, 2009a) and
with constraints from satellite kinematics (More \etal 2009). We have
used this CLF model in Section~\ref{sec:lassign} to construct our mock
group catalogues, except that we imposed that the central galaxy is
always the BHG; if a satellite luminosity was drawn from $\Phi_{\rm
  sat}(L|M)$ that was brighter than the central luminosity, drawn from
$\Phi_{\rm cen}(L|M)$, it was rejected. Later, we then switched the
luminosities of the central and that of its brightest satellite in a
fraction $\fFU$ of all groups. We now use the CLF to `predict' the
fraction $\fFU$ as function of halo mass, by making the assumption
that the luminosities of satellite galaxies are independent of the
luminosity of the central galaxy in the same halo.  In that case, the
probability that a random satellite galaxy in a halo of mass $M$ has a
luminosity smaller than that of its central is simply given by
\begin{eqnarray}\label{probLsatM}
P(L_{\rm sat}<L_{\rm cen}|M) = \phantom{screw flanders screw flanders} \\
\qquad \int_0^{\infty} 
{\int_{L_{\rm min}}^{L_{\rm cen}} \Phi_{\rm sat}(L|M) \, \rmd L \over
\int_{L_{\rm min}}^\infty \Phi_{\rm sat}(L|M) \, \rmd L} \,
\Phi_{\rm cen}(L_{\rm cen}|M)\,\rmd L_{\rm cen} \,,\nonumber
\end{eqnarray}
where the integral accounts for the scatter in the relationship
between central galaxy luminosity and halo mass, and $L_{\rm min}$ is
the minimum luminosity considered. In a halo with $N_{\rm sat}$
satellites, the probability that the central galaxy is also the 
brightest galaxy is simply $[P(L_{\rm sat}<L_{\rm cen}|M)]^{N_{\rm sat}}$.
Hence, we obtain that
\begin{equation}
\fFU(M) = 1 - \displaystyle\sum_{N_{\rm sat}=1}^{\infty}
P(N_{\rm sat}|M) \left[P(L_{\rm sat}<L_{\rm cen}|M)\right]^{N_{\rm sat}}
\label{probLsat1M}
\end{equation}
where $P(N_{\rm sat}|M)$ gives the probability that a halo of mass $M$
contains $N_{\rm sat}$ satellites, which we take to be a Poisson
distribution whose mean is given by Eqn.~(\ref{meanNsat}). The results
thus obtained from the CLF of C09 are shown as a solid line in
Fig.~\ref{probLsat1Mgrp}.  For comparison, the solid triangles with
error bars are the constraints on $\fFU(M)$ obtained from the SDSS
group catalogue as described in the previous section.  Although the
CLF predicts that $\fFU$ increases with increasing halo mass, in
qualitative agreement with the data, the values of $\fFU(M)$ inferred
from the data are much larger than those `predicted' by the CLF. Note
that the CLF prediction is for {\it all} haloes, rather than for
haloes (or groups) with $N_{\rm sat} \geq 3$, which is the richness
threshold used in our analysis of SDSS groups. However, this has
little to no impact. We have verified that our constraints on $\fFU$
do not change significantly if we use a different richness threshold.
Furthermore, from the CLF, the probability of a halo hosting fewer
than three galaxies is low for most of the mass range considered here:
$P(N_{\rm sat}<3)>0.05$ only for haloes with ${\rm log}~M<12.75$.

There are a number of possible explanations for why the CLF prediction
is not consistent with the data. First of all, we emphasize that the
CLF is not designed to `predict' $\fFU(M)$. Most of the statistics
used to constrain the CLF depend only very weakly (clustering and
lensing) or not at all (luminosity function) on $\fFU(M)$. Secondly,
it is still possible that the CLF is correct, but that the additional
assumption that the satellite luminosity is independent of the
luminosity of the central is not correct, that is, $\Phi_{\rm sat}(L_{\rm
  sat}|M,L_{\rm cen}) \ne \Phi_{\rm sat}(L_{\rm sat}|M)$. This could
come about, for example, because of galactic `cannibalism': those haloes
in which the central has recently cannibalised a bright satellite will
have an excessively bright central, and are less likely to have a
bright satellite.  It remains to be seen whether a model for
$\Phi_{\rm sat}(L_{\rm sat}|M,L_{\rm cen})$ can be found that yields a
$\fFU(M)$ in better agreement with the data, and simultaneously obeys
the CLF constraint that
\begin{equation}
\Phi_{\rm sat}(L_{\rm sat}|M) = \int\Phi_{\rm sat}(L_{\rm sat}|M,L_{\rm cen})
\, \Phi_{\rm cen}(L_{\rm cen}|M) \, \rmd L_{\rm cen}\,.
\end{equation}

Of course, it is also possible that the discrepancy reflects an actual
failure of the CLF. One possible modification, which will have little
impact on the luminosity function, clustering, galaxy-galaxy lensing,
and mock group catalogues, is a modification in the shape of
$\Phi_{\rm sat}(L|M)$ (eqn.~\ref{CLFsat}) at the bright
end. As mentioned in Section~\ref{sec:lassign}, C09 adopted a
functional form for which $\Phi_{\rm sat}(L|M) \propto
\exp\left[-(L/L_{\rm sat}^{\ast})^s\right]$ at the bright end, with
$s=2$. The exact value of $s$, though, is poorly constrained by the
data, but has a significant impact on $\fFU(M)$. This is illustrated
by the dotted and dashed curves in Fig.~\ref{probLsat1Mgrp}, which
correspond to $s=1$ and $s=3$, respectively. Clearly, decreasing $s$
increases the expectation value for the luminosity of the brightest
satellite, and hence the fraction of haloes for which the central
galaxy is not the BHG. For $s=1$ the CLF `predicts' a $\fFU(M)$ in
good agreement with the data, but only for $M\gta 5 \times 10^{13}
h^{-1} \Msun$. For less massive haloes, the value of $\fFU$ inferred
from the SDSS group catalogue is still too high compared to the CLF
prediction, but only by about $2\sigma$.

Another parameter that has a significant impact on $\fFU(M)$ is the
ratio $Q \equiv L_{\rm sat}^\ast/L_{\rm cen}$. Motivated by the CLF
inferred from the SDSS group catalogue by Yang \etal (2008), C09
adopted $Q=0.562$, with no dependence on halo mass.  However, Hansen
\etal (2009), using the maxBCG group catalogue of Koester \etal
(2007), found that $Q$ depends on halo mass and may be as small as
$\sim 0.15$ for $M \sim 10^{14} h^{-1} \Msun$.  In order for the CLF
to yield $\fFU(M)$ in rough agreement with our findings, though, we
need a value of $Q$ that is {\it larger} than $0.562$, (i.e., for
$s=2$ we require $Q \sim 0.7$).  These results warrant a more thorough
investigation into the exact values of $s$ and $Q$ as function of halo
mass.

Finally, we have verified that the CLF predictions for $\fFU(M)$ 
do not depend significantly on our choice for the lower luminosity
cut-off, $L_{\rm min}$.

\subsection{Comparison with Semi-Analytical Models}
\label{sec:compSAM}

We now compare our results to the predictions of two semi-analytical
models (SAMs) of galaxy formation; the MORGANA model of Monaco,
Fontanot \& Taffoni (2007), as updated by Lo Faro \etal (2009), and the
SAM of Croton \etal (2006).

Both SAMs adopt flat $\Lambda$CDM cosmologies, albeit with slightly
different values for the cosmological parameters\footnote{We do not
believe that these small differences will have a significant impact on
the predictions of $\fFU(M)$.}. Although both SAMs include treatments
of cooling, star formation, feedback from supernovae and active
galactic nuclei, mergers, starbursts and disk instabilities, the
actual implementations of these physical processes are substantially
different (see the original papers for details).  Yet, as shown in
Fig.~\ref{probLsat1Mgrp} (open symbols), they predict fairly similar
values for $\fFU(M)$. \footnote{As in our analysis of SDSS galaxy
groups, the predictions of $\fFU(M)$ from the SAMs are the fractions
of haloes in which a satellite galaxy is the most luminous galaxy.
The SAMs' predictions are almost exactly same for the fractions of
haloes in which a satellite is the most massive galaxy (in terms of
stellar mass).}  In qualitative agreement with the data, both SAMs
predict that $\fFU$ increases with halo mass.
In addition, from studying the satellite BHGs in the MORGANA model, we
find that the majority of the more massive satellites represent a
recently (since $z\sim0.2$) accreted population, in most cases linked
to the last major merger experienced by the host halo.  In other
words, infalling massive satellites could be a contributing cause of a
nonzero $\fFU$.

As with the halo occupation statistics, the predictions of both models
are significantly lower than the data.  In both SAMs, satellite
galaxies are `strangulated' after being accreted by their host
galaxies, so that they do not participate in the cooling flow of their
parent halo. However, it has been pointed out that the standard,
instantaneous implementation of this strangulation causes an
over-quenching of satellite galaxies (e.g., Baldry \etal 2006;
Weinmann \etal 2006; Kimm \etal 2009). It has been suggested that this
over-quenching problem can be avoided by adopting a longer time-scale
for strangulation (e.g., Kang \& van den Bosch 2008; Font \etal 2008;
Weinmann \etal 2009). This is likely to allow satellite galaxies to
continue forming stars for some period, which will increase their
stellar mass, and thus also $\fFU$. The effect, though, is likely to
be small. Another effect that may cause the SAMs to underpredict
$\fFU$ is the fact that they often adopt dynamical friction
time-scales that are too short (Boylan-Kolchin et al. 2008; Wetzel \&
White 2010). This results in (massive) satellites being accreted too
rapidly, and hence in an underestimate of $\fFU$.  Finally, the
models don't take proper account of stellar mass stripping due to
tidal forces, which will make satellites {\it less} massive. As
emphasized in a number of recent papers, this tidal stripping of
satellite galaxies is an important ingredient of galaxy formation
(Monaco \etal 2006; White et al. 2007; Conroy, Ho \& White 2007;
Conroy, Wechsler \& Kravtsov 2007; Kang \& van den Bosch 2008; Yang,
Mo \& van den Bosch 2009b; Pasquali \etal 2010). It remains to be
seen to what extent these additions and/or modifications of the
semi-analytical models impact $\fFU(M)$. For the moment, we conclude
that the $\fFU(M)$ inferred from our analysis of SDSS galaxy groups is
uncomfortably high compared to predictions from both halo occupation
statistics and from semi-analytical models of galaxy formation.

\subsection{Implication for Satellite Kinematics}
\label{sec:satkin}

Studies that attempt to infer the masses of dark matter haloes using
the kinematics of satellite galaxies always assume that the CGP is
valid (e.g., Zaritsky \etal 1993; McKay \etal 2002; van den Bosch
\etal 2004; More \etal 2009). Since a typical central only has a few
satellites, one normally stacks many centrals together in order to
obtain sufficient signal-to-noise to measure a reliable satellite
velocity dispersion, $\sigma_{\rm sat}$. With sufficiently large
galaxy redshift surveys, one can measure $\sigma_{\rm sat}$ as a
function of the luminosity (or stellar mass) of the central galaxies
(i.e., one stacks centrals in narrow bins in luminosity or stellar
mass).

Consider a halo with $N_{\rm sat}$ satellites with velocities $v_{{\rm
    sat},i}$ with respect to halo centre, and let the central be
stationary at the centre of the halo (i.e., $v_{\rm cen}=0$). The
satellite velocity dispersion measured with respect to the central is:
\begin{equation}  
\sigma_{\rm true}^2 = {1\over N_{\rm sat}}
\displaystyle\sum_{i=1}^{N_{\rm sat}} v_{{\rm sat},i}^2\,.
\end{equation}
Without loss of generality, assume that satellite number $1$ is
misidentified to be the central. The {\it measured} velocity
dispersion in this case will be:
\begin{equation}
\sigma_{\rm meas}^2 = {1\over N_{\rm sat}} \left[
\displaystyle\sum_{i=2}^{N_{\rm sat}} (v_{{\rm sat},i} - v_{\rm sat, 1})^2 
+ (v_{\rm cen}-v_{\rm sat,1})^2\right]
\end{equation}
The first term sums over the remaining $N_{\rm sat}-1$ true
satellites, while the last term is the contribution from the true
central.  Using that $\langle v_{{\rm sat},i} \rangle = 0$, and that
$\langle v_{\rm cen} \rangle = \langle v_{\rm cen}^2 \rangle = 0$, it
is easy to show that the velocity dispersion measured from a stack of
many systems, each with $N_{\rm sat}$ satellites, is equal to
\begin{equation}\label{sigeffect}
{\sigma_{\rm meas}^2 \over \sigma_{\rm true}^2} = 1 + 
\fFU\,(1 - N^{-1}_{\rm sat})\,,
\end{equation}
where $\fFU$ is the fraction of systems in which the central is
misidentified. Note that when the number of satellites per central is
one, the measured velocity dispersion is identical to the true one,
independent of $\fFU$, that is, the fact that some satellites are
misidentified as centrals has no impact on the inferred satellite
kinematics.\footnote{This statement ignores the fact that a non-zero
$\fFU$ also impacts the interpretation of the stacking parameter
(i.e., the luminosity of the alleged `centrals') and may result in
more interlopers.}
However, in the limit $N_{\rm sat} \rightarrow \infty$, one has that
$\sigma_{\rm meas} = \sqrt{1+\fFU} \sigma_{\rm true}$. Since the
inferred halo mass $M \propto \sigma_{\rm sat}^{3}$, this implies that
the mass of the halo will be overestimated by a factor
$(1+\fFU)^{3/2}$. Note that Eqn.~(\ref{sigeffect}) is strictly valid
only for stacks of systems that all have the same number of
satellites. In reality, however, there will be scatter in the number
of satellites per host in the stack. In what follows we assume that we
do not make a large error if we adopt Eqn.~(\ref{sigeffect}) but with
$N_{\rm sat}$ replaced by the {\it average} number of satellites per
central, $\langle N_{\rm sat} \rangle$.
\begin{figure}
\includegraphics[width=\hsize]{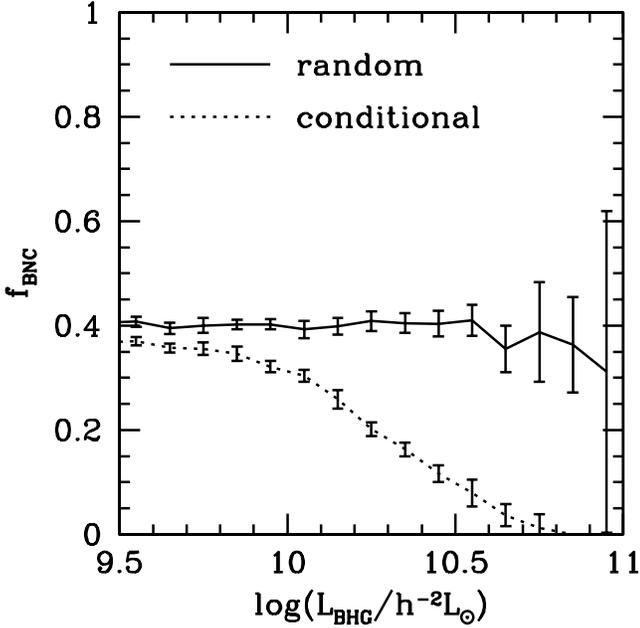} 
\caption{Fraction $\fFU$ of haloes in which the brightest galaxy is
 not the central one, as a function of the luminosity of the brightest
 halo galaxy.  The solid line shows $\fFU(L_{\rm BHG})$ for the
 `random' mocks, and the dotted line shows the fraction for the
 `conditional' mocks, in which whether a halo has a satellite brighter
 than the central is conditional on the luminosity of the central
 galaxy.  The luminosities correspond to $r$-band absolute magnitudes,
 such that ${\rm log}\,L=9.5$ corresponds to $M_r\approx-19$, and
 ${\rm log}\,L=10.3$ to $M_r\approx-21$, etc.}
\label{fig:difmocks}
\end{figure}

For relatively faint centrals, $\langle N_{\rm sat} \rangle \simeq 1$,
and $\sigma_{\rm meas} \simeq \sigma_{\rm true}$ independent of the
value of $\fFU$. However, at the bright end $\langle N_{\rm sat}
\rangle$ becomes substantially larger than unity, and a non-zero
$\fFU$ may cause a significant overestimate of the true $\sigma_{\rm
  sat}$. For example, in their analysis of satellite kinematics, More
\etal (2009) have $\langle N_{\rm sat} \rangle \sim 10$ for their
brightest host bins. If $\fFU$ for these bins is comparable to
$\fFU(M)$ at the massive end, (i.e., $\fFU \sim 0.4$), their inferred
halo masses around bright host galaxies will be overestimated by a
factor $\sim 1.6$. 

However, since the halo mass luminosity relation for central galaxies
is not one-to-one, it is not trivial to infer $\fFU$ for a certain bin
in host galaxy luminosity, $L_{\rm host}$, from $\fFU(M)$. In fact,
one can construct mocks in which $\fFU(L_{\rm host})$ is very small at
the bright end, even when $\fFU(M)$ is large at the massive end, by
making the criterion for a halo having a satellite brighter than the
central conditional on the luminosity of the central galaxy.  We
illustrate a particular case of this in Fig.~\ref{fig:difmocks}, where
we show $\fFU(L_{\rm BHG})$ for two types of mocks that have identical
$\fFU(M) = 0.4$. These are constructed as follows.  We start by
populating the dark matter haloes in our $100 h^{-1} \Mpc$ simulation
box described in Section~\ref{sec:mocks} with galaxies of different
luminosities using the CLF of C09 (see Section~\ref{sec:lassign} for
details).  If the luminosity of a satellite is brighter than that of
its central, a new luminosity is drawn until it is fainter than that
of the central, so that $\fFU=0$.  For the first set of mocks, we
switch the luminosities of the central and that of its brightest
satellite in a random fraction of 40 percent of all haloes, and we
measure the resulting $\fFU(L_{\rm BHG})$. We refer to these as the
`random' mocks. The mean and scatter obtained from ten realizations
are shown as the solid line in Fig.~\ref{fig:difmocks}. As expected,
$\fFU(L_{\rm BHG}) \simeq 0.4$, independent of $L_{\rm BHG}$.  For the
second set of mocks, we again start from the mocks with $\fFU=0$, but
we now switch the luminosities of brightest satellite and central in
those haloes that meet the criterion
\begin{equation}
\int_0^{L_{\rm cen}} \Phi_{\rm cen}(L|M) \rmd L < 0.4\,,
\end{equation}
that is, in those haloes in which the luminosity of the central falls
in the lower 40 percentile of its distribution, $\Phi_{\rm
cen}(L|M)$. We refer to these as the `conditional' mocks, and similar
to the `random' mocks they thus have $\fFU(M) = 0.4$. Yet, their
$\fFU(L_{\rm BHG})$, indicated by the dotted curve in
Fig.~\ref{fig:difmocks}, are very different: they decrease from $\sim
0.4$ at the faint end to almost zero at the bright end. Hence, if the
probability that a central is not the BHG is conditional on the
luminosity of the central, which seems to be a reasonable assumption,
then even a relatively large $\fFU(M)$ may have negligible impact on
the halo masses inferred from satellite kinematics.

We thus conclude that a proper assessment of the impact of a non-zero
$\fFU(M)$ on the halo masses inferred from satellite kinematics
requires an independent assessment of the fraction $\fFU$ as function
of the luminosities of the host galaxies.

\section{Conclusions}
\label{sec:concl}

It is generally assumed that the central galaxy in a dark matter halo,
that is, the galaxy with the lowest specific potential energy, is also
the brightest halo galaxy (BHG) and that it resides at rest at the
centre of the dark matter potential well. This central galaxy paradigm
(CGP) is an essential assumption made in various fields of
astronomical research (e.g., satellite kinematics, gravitational
lensing, both weak and strong, halo occupation modelling).

In this paper, we have used a large galaxy group catalogue, constructed
from the SDSS DR4 by Yang \etal (2007), in order to test the validity
of the CGP. For each group we compute two statistics, $\calR$ and
$\calS$, which quantify the offsets of the line-of-sight velocities
and projected positions of brightest group galaxies relative to the
other group members. By comparing the cumulative distributions of
$\vert\calR\vert$ and $\vert\calS\vert$ to those obtained from
detailed mock group catalogues, we have tested the null-hypothesis,
$\calH_0$, that the CGP is correct; hypothesis $\calH_1$, according to
which central galaxies are BHGs but have a non-zero velocity with
respect to the halo centre, parameterized by a non-zero velocity bias,
$b_{\rm vel}$; and hypothesis $\calH_2$, according to which central
galaxies reside at rest at the centre of the halo's potential well,
but are not the BHGs in a fraction $\fFU$ of all haloes.

In agreement with vdB05, who only used the $\calR$ statistic, we show
that the null-hypothesis is strongly ruled out. However, contrary to
vdB05, who argued that the data are consistent with a non-zero $b_{\rm
  vel}$, we show that $\calH_1$ fails to {\it simultaneously} match
the $\vert\calR\vert$ and $\vert\calS\vert$
distributions\footnote{Since vdB05 did not consider the spatial
  offsets (i.e., the $\calS$ statistic), they were unable to notice
  this problem for the $\calH_1$ hypothesis.}. Rather, we have shown
that the data is consistent with hypothesis $\calH_2$, indicating that
BHGs are not central galaxies in a non-negligible fraction of all
haloes; in particular, we find that $\fFU$ increases from $\sim 0.25$
in low mass haloes ($10^{12} h^{-1} \Msun \leq M \lta 2 \times 10^{13}
h^{-1}\Msun$) to $\sim 0.4$ in massive haloes ($M \ga 5 \times
10^{13} h^{-1} \Msun$).

Considering combined models, in which BHGs are not central galaxies in
a non-zero fraction $\fFU$ of all haloes, {\it and} central galaxies
have a non-zero value for their velocity bias $b_{\rm vel}$, we find
that the data can at most accommodate a small amount of velocity bias
($b_{\rm vel} \lta 0.2$). We emphasize, though, that a non-zero
velocity bias is not {\it required} by the data (see vdB05 for a
discussion of possible physical explanations for a non-zero $b_{\rm
vel}$).

Our main result is that the fraction $\fFU$ is significant and
increases with halo mass.  Since some authors have assumed that the
most luminous (or most massive) galaxy in a system is the central
galaxy, our result that $\fFU$ is somewhat large in galaxy groups and
larger ($43_{-3}^{+2}\%$) in more massive clusters may seem
surprising.  Nevertheless, our results are consistent with other
studies.  For example, von der Linden et al. (2007) find that in 343
of their 625 clusters ($\approx55\%$), the identified BCG is not the
`mean' galaxy (which lies at the centre of the cluster's density
field).  Coziol et al. (2009) find that in about half of their 452
clusters, the BCGs have a median peculiar velocity greater than one
third of their clusters' velocity dispersion.  Finally, in the Local
Group, although the Milky Way and Andromeda both have their own
systems of satellites, they could be considered part of a single
group, in which Andromeda would be identified as the BHG (van den
Bergh 1999).  The two galaxies are expected to eventually merge,
although neither is clearly the `central' galaxy.

In order to put our constraints on $\fFU(M)$ in perspective, we have
compared them to predictions from semi-analytical models (SAMs) for
galaxy formation and from halo occupation statistics. Both the SAM of
Croton \etal (2006) and that of Lo Faro \etal (2009) predict $0.1 \lta
\fFU \lta 0.2$ in the halo mass range of $10^{13} h^{-1} \Msun \lta M
\lta 10^{15} h^{-1} \Msun$, significantly lower than the $\fFU(M)$
inferred here from our SDSS galaxy group catalogue. 

We can also use the CLF model of C09 to predict $\fFU(M)$, if we
assume that the luminosity of a satellite galaxy is independent of the
luminosity of its corresponding central.  Although the C09 halo
occupation model accurately matches the SDSS $r$-band luminosity
function of Blanton \etal (2003), the projected two-point correlation
functions of Wang \etal (2007), and the galaxy-galaxy lensing data of
Mandelbaum \etal (2006), this model also predicts a $\fFU(M)$ that is
far too low. We have shown that one can increase the predicted
$\fFU(M)$ using some small modifications of the CLF, but it remains to
be seen how these modifications impact the clustering and lensing
predictions. For completeness, we emphasize that halo occupation
models based on the abundance-matching method (e.g., Vale \& Ostriker
2004; Conroy, Wechsler \& Kravtsov 2006, 2007; Shankar \etal 2006; Guo
\etal 2010; Moster \etal 2010) `predict' that $\fFU=0$, by
construction, unless they assume a non-zero scatter in the relation
between central galaxy luminosity and halo mass.  Typically a larger
amount of scatter will imply a larger $\fFU$.  It remains to be seen
whether the amount of scatter required to match the $\fFU(M)$ inferred
here is consistent with independent constraints, such as those
obtained from satellite kinematics (More \etal 2009). All in all, we
conclude that the constraints on $\fFU(M)$ obtained in this paper are
uncomfortably high compared to predictions from galaxy formation
models and halo occupation statistics. One possible explanation
  may be that dark matter haloes have substructure, something that we
  have ignored in our analysis. Although simple tests suggest that the
  correlations in the phase-space parameters of satellite galaxies due
  to substructure are not strong enough to significantly impact our
  conclusions, we caution that more detailed studies are required to
  confirm this.

Our results have important implications for various areas in
astrophysics. In particular, we have shown that a non-zero $\fFU$ may
cause an overestimate of halo masses inferred from satellite
kinematics.  Although the effect is expected to be negligible for
faint host galaxies, because they typically only have of the order of
one satellite per host, the overestimate can be significant at the
bright end. The exact impact, though, depends on the fraction $\fFU$
as function of the host luminosity, which may be very different from
that as a function of halo mass. For example, if BHGs are not centrals
in the fraction $\fFU(M)$ of haloes of mass $M$ that host the faintest
centrals, then $\fFU(L)$ drops towards zero at the bright end, and the
impact on satellite kinematics is negligible.

Additional methods and analyses that may be affected by a non-zero
$\fFU$ are the following:
\begin{itemize}

\item The inference of halo masses from weak gravitational lensing.
  Similar to satellite kinematics, weak lensing studies often rely on
  stacking the lensing signals of many clusters and groups binned by
  mass-correlated observables such as richness and luminosity.  When
  interpreting such data, it is generally assumed that BHGs coincide
  with the centres of the dark matter haloes (e.g., Mandelbaum \etal
  2006; Johnston \etal 2007; Sheldon \etal 2009a,b; Corless \& King
  2009; C09).  Since satellite galaxies yield a different lensing
  signal than centrals, at least on small to intermediate scales (see
  e.g., Yang \etal 2006), a non-zero $\fFU$ may have a significant
  impact on the lensing signal (Johnston \etal 2007), by resulting in
  underestimates of the richnesses and lensing profiles, for example.
  C09 used the CLF described in Section~\ref{sec:lassign} to model the
  galaxy-galaxy lensing signal of Mandelbaum \etal (2006). As shown in
  Section~\ref{sec:compCLF}, this CLF predicts $\fFU(M)$ significantly
  lower than inferred here.  It remains to be seen to what extent this
  may effect their interpretation of the lensing data.

\item Analyses of the power spectrum of luminous red galaxies (Tegmark
  \etal 2006; Reid \etal 2010), which can be used to constrain
  cosmological parameters. In tests with mock galaxy catalogues, Reid
  \etal (2010) find that a fraction $\fFU$ of 0.2-0.4 results in an
  angle-averaged LRG power spectrum that is damped by $2-4\%$ on
  scales of $k \sim 0.1 h \Mpc^{-1}$, and the effect is larger at
  larger $k$.  However, when analyzing the observed power spectrum
  with these mock catalogues, the effect of $\fFU$ on the recovered
  cosmological parameters is relatively small.

\item Measurements of the radial number density distribution of
  satellite galaxies, $n_{\rm sat}(r|M)$, in haloes of mass
  $M$. Several studies that have measured $n_{\rm sat}(r|M)$ using
  groups and clusters have assumed that the halo centre coincides with
  the location of the BHG (e.g., Carlberg, Yee \& Ellingson 1997;
  Collister \& Lahav 2005; Hansen \etal 2005; Yang \etal 2005b). If,
  instead, the BHG is a satellite galaxy, this will result in an
  underestimate of the concentration of $n_{\rm sat}(r)$. Hence, a
  non-zero $\fFU(M)$ may cast doubt on the claim, made by several of
  these studies, that satellite galaxies are less centrally
  concentrated than the dark matter.  On the other hand, Lin, Mohr \&
  Stanford (2004) used the centre of the X-ray emission as the centre
  of the dark matter halo, rather than the location of the BHG, and
  came to a similar conclusion.

\item Comparisons of the properties of central and satellite galaxies.  A
  number of recent studies have used galaxy group catalogues to split
  the galaxy population into centrals and satellites, and to compare
  their properties (e.g., Weinmann \etal 2006, 2009; Skibba \etal
  2007; van den Bosch \etal 2008; Pasquali \etal 2009, 2010; Hansen
  \etal 2009; Kimm \etal 2009; Skibba 2009; Guo \etal 2009).  Since
  all of these studies have assumed the brightest group galaxy to be the
  central galaxy, they are likely to have underestimated the true
  differences (i.e., a non-zero $\fFU$ blurs the actual differences 
  between centrals and satellites).

\end{itemize}

\section*{Acknowledgements}

We thank Darren Croton, for making the results from his semi-analytical
model available to us in electronic format, and Eric Bell, Kris
Blindert, Romeel Dav\'{e}, Xi Kang, Tod Lauer, Alexie Leauthaud, 
Pierluigi Monaco, Beth Reid, Maria Pereira, Simone Weinmann, and Ann
Zabludoff for valuable discussions about our results and their
implications. Some of the calculations were carried out on the PIA
cluster of the Max-Planck-Institute f\"{u}r Astronomie at the
Rechenzentrum Garching.

\label{lastpage}

\end{document}